\newcommand{\mnras}{MNRAS} \newcommand{\apj}{ApJ}
\newcommand{\nat}{Nat} \newcommand{\apjl}{ApJL}
\newcommand{\apjs}{ApJS} 
\newcommand{\aap}{A\&A} \newcommand{\aaps}{A\&AS}
 \newcommand{\aj}{AJ}
 
\newcommand{\pasp}{PASP} 

\documentclass[useAMS,usenatbib]{mn2e} \usepackage{amssymb}
\usepackage{epsfig}  \usepackage{rotating} \usepackage{graphicx}
\usepackage{longtable}  \usepackage{supertabular}
\usepackage{subfigure} \usepackage{ amssymb } \usepackage{fmtcount}

 \def\Msun{\hbox{$\rm\thinspace M_{\odot}$}}
\def\asec{$^{\prime\prime}$}

\begin{document}
\hsize=6truein

\title[The galaxy size-mass relation at $z\simeq1.4$]{The sizes,
  masses and specific star-formation rates of  massive galaxies at
  $\bmath{ 1.3<z<1.5}$: strong evidence in favour of evolution via
  minor mergers} \author[R. J. McLure et
  al.]{R. J. McLure$^{1}$\thanks{Email: rjm@roe.ac.uk},
  H.J. Pearce$^{1}$, J.S. Dunlop$^{1}$, M. Cirasuolo$^{1}$,
  E. Curtis-Lake$^{1}$, \and V.A. Bruce$^{1}$, K.I. Caputi$^{1,2}$,
  O. Almaini$^{3}$, D.G. Bonfield$^{4}$, E. J. Bradshaw$^{3}$, \and F. Buitrago$^{1}$
  R. Chuter$^{3}$, S. Foucaud$^{5}$, W. G. Hartley$^{3}$, M.J. Jarvis$^{4}$\\ $^{1}$SUPA\thanks{Scottish
    Universities Physics Alliance} Institute for Astronomy, University
  of Edinburgh, Royal Observatory, Edinburgh EH9 3HJ\\ $^{2}$Kapteyn Astronomical 
Institute, University of Groningen, P.O. Box 800, 9700 AV Groningen, The 
Netherlands\\ $^{3}$School
  of Physics and Astronomy, University of Nottingham, University Park,
  Nottingham NG7 2RD\\ 
$^{4}$Centre for Astrophysics, Science \& Technology Research Institute, University of Hertfordshire, Hatfield, Herts AL10 9AB
\\
$^{5}$Department of Earth Sciences, National
  Taiwan Normal University, Taipei 11677, Taiwan}

\maketitle

\begin{abstract}
We report the results of a comprehensive study of the relationship
between galaxy size, stellar mass and specific star-formation  rate
(sSFR) at redshifts $1.3<z<1.5$. Based on a mass complete ($M_{\star}\geq 6\times
10^{10}\Msun$), spectroscopic sample from the  UKIDSS Ultra-deep
Survey (UDS), with accurate stellar-mass measurements derived from
spectro-photometric fitting,  we find that at $z\simeq 1.4$ the
location of massive galaxies on the size-mass plane is
determined primarily by their sSFR. At this epoch we find that massive galaxies
which are passive (sSFR $\leq 0.1$ Gyr$^{-1}$) follow a tight size-mass relation, with
half-light radii a factor $f_{g}=2.4\pm{0.2}$ smaller than their local 
counterparts. Moreover, amongst the passive sub-sample we
find no evidence that the off-set from the local size-mass relation is
a function of stellar population age.  
In contrast, we find that massive star-forming galaxies at this epoch lie closer to the local late-type size-mass 
relation and are only a factor $f_{g}=1.6\pm{0.2}$ smaller than observed locally. 
Based on a sub-sample with dynamical
mass estimates, which consists of both passive and star-forming
objects, we also derive an independent estimate of $f_{g}=~2.3\pm{0.3}$ for the
typical growth in half-light radius between $z\simeq 1.4$ and the present
day. Focusing on the passive sub-sample, we conclude that to produce
the necessary evolution predominantly via major mergers would require
an unfeasible number of merger events and over populate the high-mass
end of the local stellar mass function.  In contrast, we find that a
scenario in which mass accretion is dominated by minor mergers can
comfortably produce the necessary evolution, whereby an increase in stellar mass 
of only a factor of $\simeq 2$, accompanied by size growth of a factor of $\simeq 3.5$, is required to reconcile the size-mass relation 
at $z\simeq 1.4$ with that observed locally. Finally, we note that a significant fraction ($44\%\pm12\%$)
of the passive galaxies in our sample have a disk-like morphology, providing additional evidence that 
separate physical processes are responsible for the quenching of
star-formation and morphological transformation in massive galaxies.
\end{abstract}


\section{Introduction}

Following the introduction of near-IR selected galaxy surveys nearly a
decade ago, it became clear that a substantial population of
massive, apparently evolved, galaxies was already in place by $z\simeq
1$ (e.g. Cimatti et al. 2002; Glazebrook et al. 2004), a result which was completely
unexpected within the context of contemporary models of galaxy formation
and evolution (e.g. Kauffmann \& Haehnelt 2000). 
Moreover, in line with our current understanding of galaxy ``downsizing'', it was quickly 
realised that the high-mass end of the galaxy stellar mass function was assembled much
earlier than had been expected, with studies of near-IR selected samples confirming that
$\geq 75\%$ of the local number density of $M_{\star}\geq 3\times 10^{11}\Msun$ galaxies was 
already in place by $z\simeq 1$ (e.g. Caputi et al. 2005). 

Although the latest generation of galaxy evolution models are now more
successful at reproducing the observed  number densities of high-mass
galaxies as a function of redshift (e.g. Cirasuolo et al. 2010), it
has been known  for several years that the physical properties of
high-mass galaxies at $z\geq 1$ are substantially different  from
their low-redshift counterparts. Specifically, based on HST imaging of
near-IR selected galaxies in  the Hubble Ultra Deep Field, Daddi et
al. (2005) noted that many high mass ($M_{\star}\geq 10^{11}\Msun$)  early-type
galaxies (ETGs) at $z\geq 1.5$ displayed substantially smaller
half-light radii than equivalently massive galaxies observed in the local
Universe. Over the last few years, the size evolution of massive ETGs
has been extensively discussed in the  literature and many studies
(e.g. Trujillo et al. 2007; Cimatti et al. 2008; Saracco et al. 2009; Williams et al. 2010) have
reported that massive red/passive ETGs at $1<z<3$ are a factor of
$2-5$ smaller than ETGs of the same mass in the SDSS (Shen et al. 2003). 

Despite initial concerns that results regarding the compactness of
high-redshift passive galaxies could be  affected by
surface-brightness dimming, morphological k-corrections or even
unresolved AGN components  (e.g. Daddi et al. 2005), deep,
high-resolution, near-IR imaging has conclusively demonstrated the
compactness of many massive ($M_{\star}\geq 10^{11}\Msun$) galaxies at $z\geq
1$ (e.g. Buitrago et al. 2008; Damjanov et al. 2009).  In addition, it
is now well established that, within a given redshift and stellar mass interval, early-type/passive
galaxies are more compact than late-type/star-forming galaxies 
(e.g. Zirm et al. 2007; Toft et al. 2009; Williams et al. 2010; Wuyts et al. 2011). 

In reality, the full picture is undoubtedly more complicated still,
with some studies (e.g. Mancini et al. 2010) identifying  examples of
massive $z\geq 1$ ETGs with half-light radii fully consistent with the
local size-mass relation and others (e.g. Saracco et al. 2009)
finding that the population of high-redshift ETGs is comprised of a
mixture of normal and compact galaxies. Moreover, it is still not currently
clear whether the compact nature of some massive ETGs at high
redshift  is related to the age of their stellar population
(Saracco et al. 2009), or not (Trujillo, Ferreras \& de La Rosa 2011).  Finally,
there have been concerns raised in the literature that comparing
massive high-redshift ETGs with the local galaxy population is subject
to severe progenitor bias. For example, it has been claimed that the
number densities of compact ETGs at high redshift are actually
compatible with compact cluster galaxies seen locally (Valentinuzzi et
al. 2010), although this has been disputed by Taylor et
al. (2010). Significantly, van Dokkum et al. (2010) attempted to
circumvent the progenitor bias problem by studying the size evolution
of galaxies selected to have a constant number density of $n\simeq
2\times 10^{-4}$Mpc$^{-3}$ (equivalent to $M_{\star}\simeq
3\times 10^{11}\Msun$ at $z=0$), finding strong evidence for size
evolution of the form $R_{e}\propto (1+z)^{-1.3}$.

It is thus clear that some physical
process (or processes) must be in place at high redshift which is
capable of substantially increasing the half-light radii of a significant subset of massive
high-redshift galaxies, without requiring an unphysical increase in their stellar masses.
A widely advocated solution to the problem has
been the ``dry-merger'' scenario, whereby dense high-redshift galaxies
are transformed via a series of  dissipationless (i.e. dry)
mergers (e.g. Cimatti, Nipoti \& Cassata 2012). Such dry mergers are typically envisioned to be major-mergers
(i.e. mass ratio $\geq$ 1:3) in which the galaxy half-light radius
increases in direct proportion to the accreted stellar mass.  
Although this scenario has several desirable characteristics, it also suffers from several obvious 
problems (see Nipoti et al. 2009; 2012). Principal among these is the prediction from N-body
simulations that massive $M_{\star}\geq 10^{11}\Msun$ galaxies at $z\simeq 1.5$ 
are only likely to undergo one major merger over the proceeding $\simeq 9$ Gyr 
(e.g. Hopkins et al. 2010), strongly suggesting that major mergers {\it alone} cannot 
explain the size evolution of massive high-redshift galaxies.

An alternative mechanism invokes
size growth via a sequence of minor mergers, with typical mass ratios
of 1:10. The advantage of this mechanism is that in a minor merger, 
radius growth is expected to be proportional to the square of the
accreted mass (e.g. Naab et al. 2009; Bluck et al. 2012), allowing rapid size growth
without excessive growth in stellar mass. The minor-merger scenario
has recently gained support from hydro-dynamical simulations 
(e.g. Johansson, Naab \& Ostriker 2012; Oser et al. 2012) and the observational results from
van Dokkum et al. (2010) which demonstrate that, at a constant number density of
$n\simeq 2\times 10^{-4}$Mpc$^{-3}$, the size evolution of galaxies is 
consistent with an ``inside-out'' growth scenario. Finally, it has also been suggested  (e.g. Fan et
al. 2008) that feedback from active galactic nuclei could 
expel material from the central regions of high-redshift galaxies, thereby allowing them 
to expand significantly at a constant stellar mass. However, the increase in stellar-velocity 
dispersion with redshift predicted by this model may be in conflict with the latest
available data (e.g. Trujillo et al. 2011).

In this paper we analyse the sizes, stellar masses, morphologies and
specific star-formation rates of a unique sample of massive 
galaxies from the UKIDSS Ultra-Deep Survey (UDS) with spectroscopic redshifts in the range 
$1.3<z<1.5$. Based on this sample we investigate the evolution in the galaxy size-mass
relation over a look-back time of 9 Gyr, using accurate stellar-mass
measurements derived from full spectro-photometric fitting of the
available spectra and multi-wavelength photometry. Armed with
accurate stellar-mass measurements and star-formation rates based on a
range of empirical indicators, we investigate whether any off-set from
the local size-mass scaling relations is primarily driven by
morphology,  stellar population age or specific star-formation rate.

The structure of this paper is as follows. In Section 2 we describe
the relevant spectroscopic and photometric data, before describing the 
methods adopted to estimate the stellar masses and galaxy radii in Section 3.
In Section 4 we present our main results on the
stellar mass - size relation and investigate the
dynamical mass - size relation for a small sub-sample of objects
with reliable stellar-velocity dispersion measurements.
In Section 5 we outline a plausible
evolutionary scenario which is capable of reconciling the size-mass
relation of passive galaxies at $z\simeq 1.4$ with that observed locally. In Section 6 we
present our final conclusions. Throughout this work a cosmology of
$H_0 = 70$ km s$^{-1}$Mpc$^{-1}$, $\Omega_m = 0.3$ and $\Omega_\Lambda =
0.7$ is assumed and all magnitudes are quoted in the AB
system (Oke \& Gunn 1983).

\section{The Data}
All of the galaxies studied in this work have been selected from the
UKIRT Ultra Deep Survey (UDS). The UDS is the deepest of five surveys
being undertaken with the UK Infra-Red Telescope in Hawaii, which
together comprise the UK Infrared Deep Sky Survey (UKIDSS;
\citealt{law}).  The UDS covers an area of $0.8$ sq. degrees, is
centred on RA=02:17:48, Dec=$-$05:05:57 and is currently the deepest,
large area, near-IR survey in existence. In addition to the available
UKIRT near-IR imaging, the UDS is supported by a wide array of
multi-wavelength datasets, ranging from the X-ray to radio
r\'{e}gimes (e.g. Cirasuolo et al. 2010).  In this section we briefly
summarize the properties of the UDS datasets which have been directly
exploited during the course of this study.

\subsection{Photometric data}
The latest release of the UDS, data release eight (DR8), consists of
$JHK$ imaging to $5\sigma-$depths of  $J=24.9$, $H=24.3$ and $K=24.7$
(2 arcsec-diameter apertures). Moreover, in this study we have also
made use of $Y-$band imaging of the UDS (Y=23.6; $5\sigma$) obtained
during the verification phase of the VISTA VIDEO survey (Jarvis et
al. 2012, in preparation). In addition to the available near-IR
imaging, the second  key imaging dataset in the UDS is the deep Subaru
optical imaging obtained as part of the  Subaru/XMM-Newton Deep Survey
(SXDS). The SXDS imaging (Furusawa et al. 2008) covers a total area of $1.3$ sq. degrees  to
$5\sigma$ depths of $B = 27.6$, $V = 27.2$, $R = 27.0$, $i' = 27.0$
and $z' = 26.0$ (2 arcsec-diameter apertures). The final key dataset used in this work is the Spitzer
Public Legacy Survey of the UKIDSS Ultra Deep Survey (SpUDS). The
SpUDS imaging covers the full near-IR survey area and  consists of
IRAC imaging to $5\sigma$ depths of $3.6\mu$m=23.7, $4.5\mu$m=23.3,
$5.8\mu$m=21.6, $8.0\mu$m=21.4 (3.8 arcsec-diameter apertures) and 
MIPS imaging to a $5\sigma$ depth of $24\mu$m=18.9 (15 arcsec-diameter apertures).

\subsection{Spectroscopic data}\label{follow_up}
The optical spectra analysed here were
observed on the VLT as part of the systematic spectroscopic follow-up
of the UDS obtained through the ESO large programme ESO 180.A-0776
(UDSz; P.I. O. Almaini).   Although full details of UDSz will be
presented in Almaini et al. (2012, in preparation),  below we provide
a brief description of the key elements within the context of this
study.

The UDSz programme was allocated a total of 235 hours of observations,
with 93 hours allocated for observations with the VIMOS spectrograph
(8 pointings) and 142 hours allocated for FORS2 observations (20
pointings).  The primary science driver for UDSz was to obtain
spectroscopic observations of a representative sample of $K-$band
selected galaxies ($K<23$), photometrically pre-selected to lie at
redshift $z_{phot}\geq 1$.  Consequently, the selection of primary
targets was designed to obtain spectra for a  random 1-in-6 sampling
of $K<23$ galaxies with $z_{phot} \geq 1$ (with a control sample of
$\simeq500$ galaxies with $z_{phot} < 1$). 

In order to exploit the different strengths of the two spectrographs,
bluer galaxies ($V\leq 25$) were targeted with VIMOS and redder
galaxies ($V > 25 \wedge i\leq24.5$) were targeted with FORS2. All
of the galaxies analysed in this paper were part of the spectroscopic
sample observed with FORS2 and each target received 5 hours of
on-source integration with the medium-resolution GRS$\_$300I grism
($6000$\AA$~<~\lambda~<~10000$\AA\,  with $R=660$). In total 718 ($K<23$) galaxies
were targeted with FORS2, returning 451 science-grade
redshifts.

\subsection{The K-Bright galaxy sample}
\begin{figure}
\begin{center}
\includegraphics[scale=0.7]{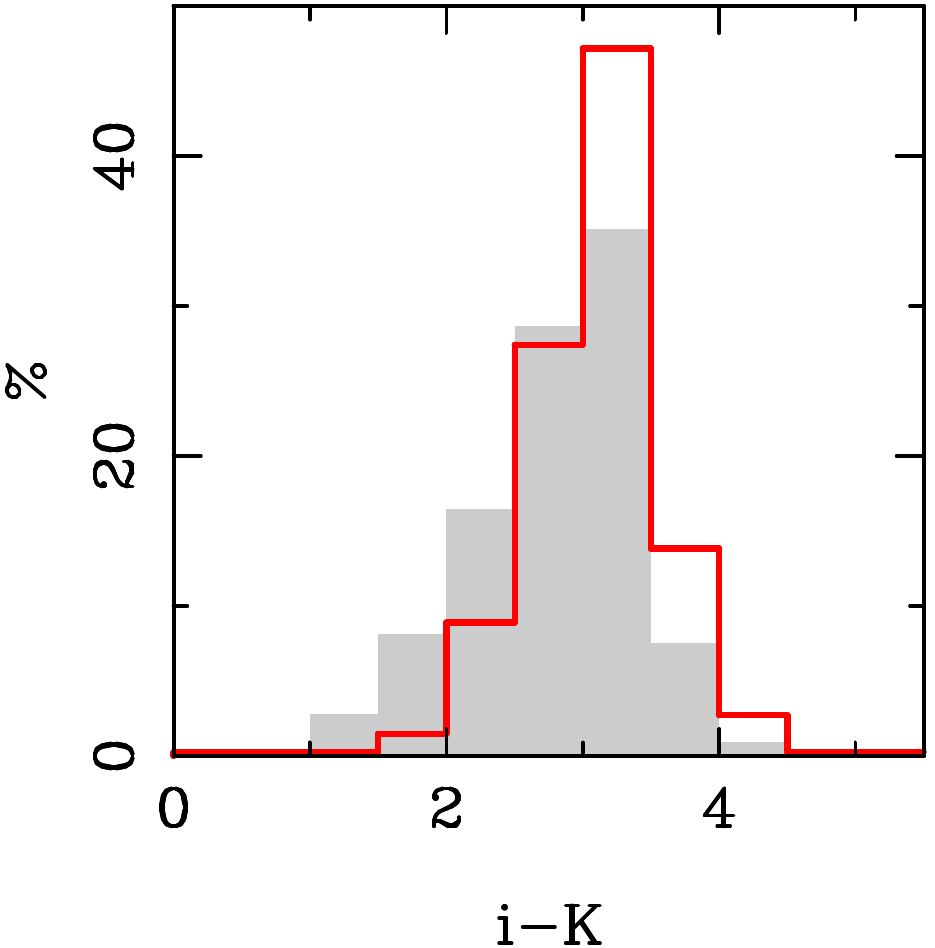}

\vspace{1.0cm}

\includegraphics[scale=0.7]{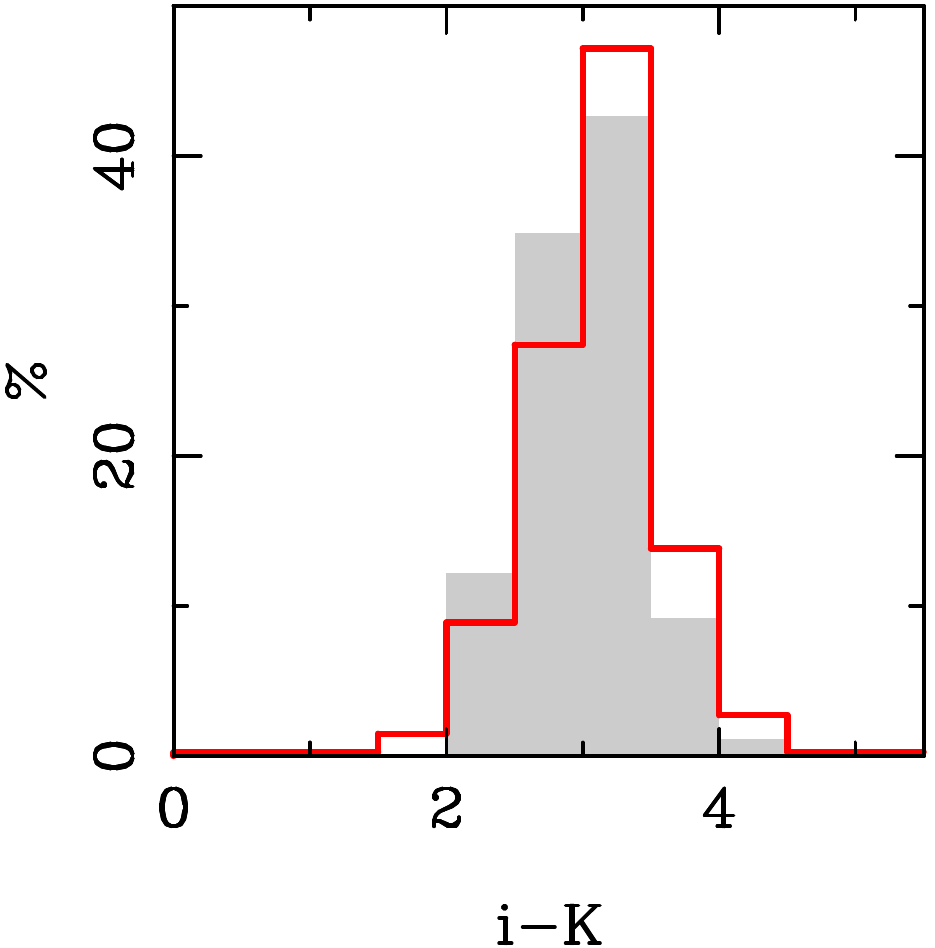}
\caption{The top panel shows the (normalised) distribution of the
  $i-K$ colours of the eighty-one galaxies in the K-Bright sample (solid red
  line) over-plotted on the $i-K$  distribution for all N=1178 galaxies
  in the UDS with $K_{tot}\leq 21.5$ and $1.3~\leq~z_{phot}~\leq~1.5$.
  The two distributions are statistically distinguishable at the $\geq
  3\sigma$ level based on a two-sample KS test, which is due to the
  fact that, by design, the FORS2 sample does not feature the tail of blue
  objects with $i-K < 2$. The bottom panel
  compares the $i-K$ distribution of the K-Bright sample to the N=970
  galaxies in the UDS with $K_{tot} \leq 21.5$, $1.3\leq
  z_{phot}\leq1.5$ and $i-K \geq 2.2$. The two distributions shown in
  the bottom panel are statistically {\it indistinguishable}
  ($p=0.22$). Given the tight relationship between $i-K$ colour and
  stellar mass for galaxies at  $z\simeq 1.4$ with $K_{tot}\leq 21.5$,
  the bottom panel demonstrates that the K-Bright sample is a
  random sample of the general $z\simeq 1.4$ galaxy  population at
  stellar masses $M_{\star}\geq 3\times 10^{10}\Msun$.}
\end{center}
\end{figure}

The sample analysed in this study consists of all eighty-one UDSz galaxies
with robust FORS2 spectroscopic redshifts (i.e. based on multiple spectral features) in the range $1.3\leq z \leq 1.5$
which are brighter than $K_{tot} \leq 21.5$ (hereafter the K-Bright sample). 
This magnitude limit is bright enough to fully exploit the
large-area coverage of the UDS and to ensure that each member of the
sample has an unambiguous spectroscopic redshift, while still being
deep enough to produce a final sample which is mass-complete to a limit of $M_{\star}=6\times10^{10}\Msun$, even for the most passive
galaxies (see Section 4). The upper redshift limit of $z=1.5$ was chosen to guarantee coverage of
the 4000\AA\, break at the red end of the FORS2 spectra and the lower redshift limit of 
$z=1.3$ was adopted to ensure that the FORS2 spectra also cover the age-sensitive
spectral breaks at 2600\AA\, and 2800\AA.

\subsubsection{A representative sample}
If the results derived from this study are to be applied to
the general massive galaxy population at $z\simeq 1.4$, it is  clearly
necessary to demonstrate that the K-Bright sample is not a highly
biased sub-set of the general galaxy population at this epoch.  In
order to investigate this issue, in the upper panel of Fig. 1 we plot
the distribution of $i-K$ colours for the K-Bright sample, compared to
that of all N=1178 galaxies in the UDS photometric redshift catalogue
of Cirasuolo et al. (2010) which satisfy the criteria: $1.3\leq z_{phot} \leq 1.5$ and 
$K_{tot}\leq21.5$. Although similar, a two-sample KS test
demonstrates that the two distributions are actually statistically
distinguishable at  the $\geq 3\sigma$ level. As can clearly be seen
from Fig. 1, unusually for a spectroscopic sample, the difference is
caused by the fact that the K-Bright sample does not feature a large
enough tail of blue ($i-K<2$) objects. This feature of the FORS2 spectroscopic sample is by design and is 
the result of deliberately targeting the
bluer members of the UDSz sample with the VIMOS spectrograph.

In the lower panel of Fig. 1 we show the $i-K$ distribution of the
K-Bright sample compared to the N=970 objects from the Cirasuolo et
al.  photometric redshift catalogue which satisfy: $1.3 \leq z_{phot} \leq 1.5$, 
$K_{tot}\leq21.5$ {\it and} $i-K\geq 2.2$. With this additional
colour criterion the two distributions are now statistically {\it indistinguishable} ($p=0.22$). For galaxies at
$z\simeq 1.4$ with $K_{tot}\leq 21.5$, insisting on a colour of
$i-K\geq 2.2$ is effectively the same as insisting that stellar mass
is $M_{\star}\geq 3\times 10^{10}\Msun$ (see Section 4). Consequently, we conclude
that the K-Bright sample is a statistically random sub-sample of the
galaxy population at $1.3 \leq z \leq 1.5$ with stellar masses $M_{\star}\geq 3\times 10^{10}\Msun$.
For information, in Table A1 of the appendix we provide the basic observational data and derived properties 
for the eighty-one members of the K-Bright galaxy sample.

\subsubsection{Sampling factor}
The UDSz spectroscopic programme
was designed to provide a random, 1-in-6 ($f_{samp}=6$), sampling of
the  $z\geq 1$ galaxy population with $K\leq 23$. However, at the $K_{tot}\leq21.5$
magnitude limit of the K-Bright sample the FORS2 component of the
UDSz spectroscopy  programme actually provides a significantly higher
sampling rate of the very high-mass end of the galaxy mass function at
$1.3<z<1.5$.  Based on the Cirasuolo et al. photometric  redshift
catalogue, and accounting for the differences in relative areal
coverage, we calculate that, compared to the underlying galaxy
population with $1.3< z_{phot} < 1.5$, $K_{tot}\leq21.5$ and $i-K\geq
2.2$, the sample factor of the K-Bright  sample is
$f_{samp}=3.7\pm{0.4}$. It is this sampling factor which is adopted
when comparing the K-Bright sample to  the local galaxy stellar mass
function in Section 5.

\section{Galaxy masses, ages and sizes}
In addition to a precise spectroscopic redshift, for each member of the
K-Bright sample there is a deep, flux calibrated, red-optical FORS2 spectrum and excellent
multi-wavelength photometry spanning the range $0.4\mu$m $< \lambda <
4.5\mu$m. Consequently, in order to take full advantage of the
available information, we have  derived stellar mass and age
measurements for each member of the K-Bright sample via simultaneous
spectro-photometric fitting of the FORS2 spectrum and
multi-wavelength photometry.

\begin{figure*}
\begin{center}
\includegraphics[width=19.0cm, height=12.0cm]{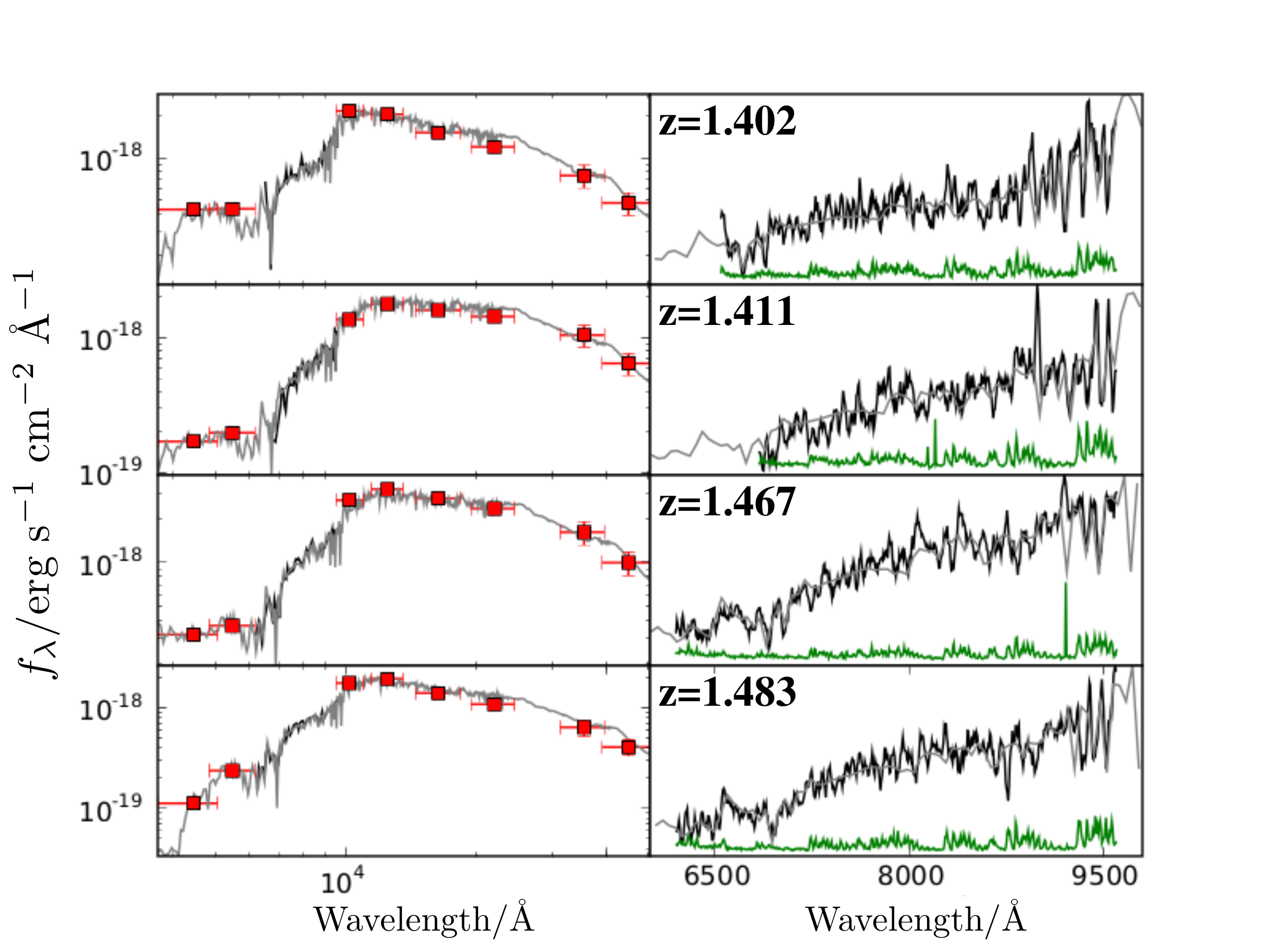}
\caption{Plots illustrating the spectro-photometric fitting process for
  four example objects from the K-Bright sample. The left-hand panels
  show the combined fits to the spectra and photometry, while the
  right-hand panels show the model fits through the
  spectra in detail. In each case the optical spectrum is shown in black, the error
  spectrum is shown in green, the photometry is in red and the
  best-fitting model is in grey. For presentation purposes the optical
  spectra have been smoothed using a 5-pixel boxcar.}
\label{tcm_fits_plot}
\end{center}
\end{figure*}

\subsection{Spectro-photometric fitting}\label{spec_phot_fit}
For the purposes of the spectro-photometric fitting we adopt both the
\cite{bc03} (hereafter BC03) and the Charlot \& Bruzual (2007)\footnote{St\'ephane Charlot, private communication}
(hereafter CB07) stellar
population synthesis models. The CB07 models were chosen because they
include an updated treatment of thermally pulsating asymptotic giant
branch (TP-AGB) stars which could, at least potentially, have a large
influence at rest-frame near-IR wavelengths for stellar populations
with ages of $0.3 \leq t \leq 2$ Gyr (e.g. Maraston 2005). However,
given that the influence of TP-AGB stars remains controversial
(e.g. Kriek et al. 2010), it was decided to  investigate the range of
stellar masses and ages returned by both the BC03 and CB07 stellar
population synthesis models.

Moreover, as galaxies are complex systems, potentially consisting of
several stellar sub-systems from different generations of
star-formation, in some cases it is necessary to invoke models with a variety
of star-formation histories  in order to obtain good fits to the
data. To account for this, the data for each member of the K-Bright
sample was fitted using an extensive set of models encompassing a range
of different star-formation histories,  metallicities and dust attenuations.

\subsubsection{Exponentially decaying star-formation rate models}
The data for each member of the K-Bright sample was initially fitted using BC03 and
CB07 models with exponentially decaying star-formation rates (SFR);
so-called ``$\tau-$models''. The $\tau-$models are parameterized such
that  SFR $ \propto \exp(-t/\tau)$, where $t$ is the time since star
formation began and $\tau$ is the $e-$folding time.  A wide range of
$e-$folding times were considered, with the extreme values of $\tau =
10$ Gyr and $\tau = 0.1$ Gyr approximating constant star-formation
and instantaneous burst models respectively. As listed in Table 1,  in
addition to $e-$folding time, a wide range of ages, metallicities and
dust attenuations were also explored. All models employed in the
spectro-photometric fitting were based on a \cite{chab} IMF and assumed the
Calzetti et al. (2000) dust attenuation curve.

\subsubsection{Double burst models}
Although the $\tau-$models effectively describe a range of different
star-formation histories,  they do not adequately describe the
situation whereby the majority of a galaxy's stellar mass is  produced
rapidly at high redshift (and dominates rest-frame near-IR
wavelengths) but a low-level of  more recent star-formation dominates
the observed flux at rest-frame UV/optical wavelengths.  In order to
account for this possibility, the data for each member of the K-Bright
sample was fitted using  so-called ``double burst models'' (hereafter DB
models). The DB models consist of two sequential  instantaneous bursts
(simple stellar populations), with one component required to be ``old''
(age $\geq 0.5$ Gyrs)  and one component required to be ``young'' (age
$<0.5$ Gyrs).

The DB models were constructed by normalising the flux of the younger
stellar population to match that of the older population at a
rest-frame wavelength of 5000\AA. In this way, when the components 
were combined, the two populations naturally
accounted for different fractions of the total galaxy mass.  The mass
fraction of the young component was allowed to vary between 0 and 1,
with a fine grid of values adopted at  low-mass fractions due to the
substantial impact  on the rest-frame UV/optical from even small
amounts of recent star-formation. Finally, dust reddening  was applied
to the combined young+old stellar population using the Calzetti (2000)
dust attenuation law.  In addition to providing improved
flexibility, the DB models offer the  advantage of providing an
estimate of the likely maximal stellar mass and age for each
member of the K-Bright sample.

\begin{table}
\caption{The parameter space of stellar population models used in the
  fitting of the K-Bright sample, where $t_{H}$ corresponds to the age of
the Universe at a given redshift.}
\begin{center}
\begin{tabular}{lr}
\hline \hline $\tau-$model parameters:&\\ &\\ 

$e-$folding time:  &$0.1$ Gyr $ <\tau<10$ Gyr \\  
Age:               &0.01 Gyr $<t<t_{H}$ \\  
Metallicity:       & $0.2{\rm Z}_\odot<{\rm Z}<2.5 {\rm Z}_\odot$\\  
Reddening:         & $0.0<A_{V}<3.0$ mags\\ & \\

\hline 
DB model parameters:&\\ &\\

Old comp metallicity:         & $0.2{\rm Z}_\odot, {\rm Z}_\odot$  \\  Old comp age:
& $0.5$ Gyr $<t_{oc}<t_{H}$ \\  Young comp metallicity:       &
$0.2{\rm Z}_\odot < {\rm Z} < 2.5 {\rm Z}_\odot$\\  
Young comp age:               & $0.01$ Gyr $< t_{yc} < 0.5$ Gyr \\  
Young comp mass fraction:     & $0.0 < m_{yc} < 1.0$\\  
Reddening:                   & $0.0<A_{V}<3.0$ mags \\ \hline \hline
\end{tabular}
\end{center}
\end{table}

\begin{figure*}
\begin{center}
\includegraphics[width=17.5cm, height=4.5cm]{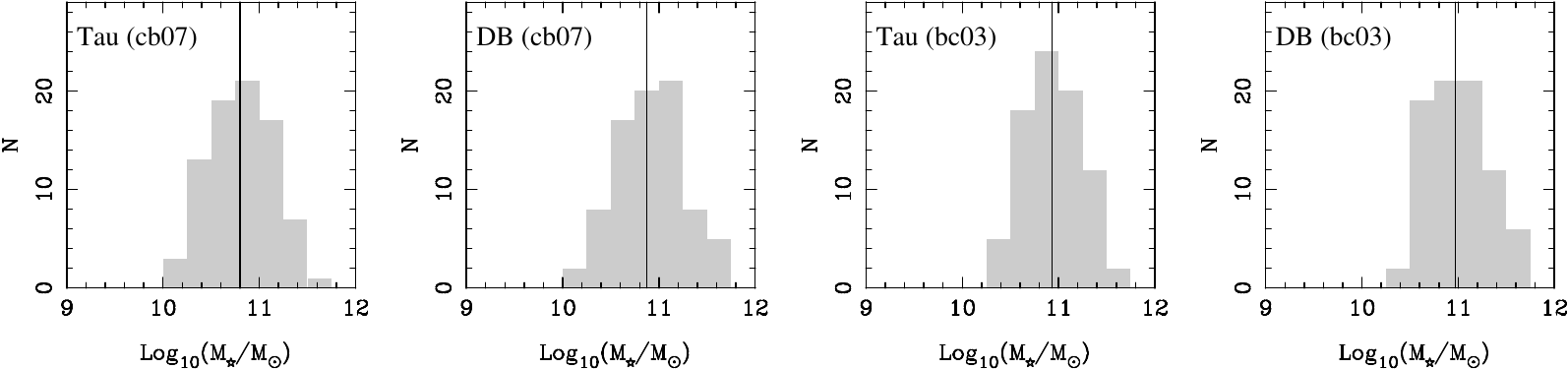}
\caption{An illustration of the different stellar-mass distributions
  for the K-Bright galaxy sample obtained via spectro-photometric
  fitting based on different stellar population models and
  star-formation histories. The stellar mass distributions are ordered
  (left-to-right) by increasing median stellar mass (vertical solid
  lines) and correspond to $\tau-$models and double-burst models
  based on the CB07 stellar population models followed by the
  equivalent models based on the BC03 stellar populations models. From
  left-to-right the median stellar mass of the sample for each
  distribution are as follows:  $6.3\times 10^{10}\Msun$, $7.4\times
  10^{10}\Msun$, $8.5\times 10^{10}\Msun$ and $9.3\times
  10^{10}\Msun$. For the  purposes of the analysis performed in this
  paper the stellar-mass measurements derived from the BC03
  $\tau-$models  have been adopted. However, as illustrated here, the difference in 
median mass between the
extreme combinations of template library/star-formation history only differ
at the $0.17$ dex level, insufficient to affect any of the science
results derived in the latter sections of the paper.}
\end{center}
\end{figure*}

\subsubsection{Model fitting procedure}

For each member of the K-Bright sample, the stellar population models
were initially shifted to the appropriate redshift and the
corresponding FORS2 spectrum was binned-up to match the resolution of
the stellar population model (typically 10\AA\, per pixel). The
stellar population models were then simultaneously fitted to the spectra
and the  accompanying $BVYJHK$ and IRAC $3.6\mu$m \& 4.5$\mu m$
photometry. During the fitting process the FORS2 spectral data in the
wavelength range 6200-9600\AA\, were included in the fit, but the
wavelength range 9600-10000\AA\, was excluded to ensure  that bad
sky-line residuals did not dominate the fit. The photometric
data covering the same  wavelength range as the spectra ($Riz$) was 
not included in the fitting process, simply because those photometric
bands  had already been used to flux normalize the spectra and
photometric datasets.

The photometric broad-band magnitudes were converted to flux densities
at the effective  wavelength of each filter and the corresponding
model predictions were calculated by integrating the models over the
appropriate filter profiles. Based on the parameter space outlined in
Table 1, each galaxy was fitted using a grid of up to $\sim5 \times
10^5$ models from both the BC03 and CB07 libraries. The best-fitting
model was determined using $\chi^2$ minimization, with the photometry 
weighted using the appropriate photometric errors and the spectral data weighted using the FORS2 error spectra.

In order to prevent the fitting procedure from being dominated by the spectral data two key steps were necessary.
Firstly, it was necessary to mask areas of the spectra which the spectral synthesis models 
were not capable of reproducing (principally the [OII] emission line). Secondly, it was necessary to ensure that the 
FORS2 error spectra actually accounted for all the inherent uncertainties present in the data (both random and systematic). 
The final reduced FORS2 spectrum for each object typically consists of 5 hours of on-source integration time, comprised of a stack of 
twenty-four individual exposures, each of 12.5 minutes integration. 
Consequently, at each spectral pixel it was possible to use these independent flux measurements to 
calculate the sigma ($\sigma_{f}$) which fully accounted for the variation in measured flux due to 
shot noise, varying observing conditions and systematic errors such as imperfect sky-line subtraction. 
By weighting the final FORS2 spectra by the standard error (i.e. $\sigma_{f}/\sqrt{24}$), it was possible to perform a 
combined $\chi^{2}$ fit to the spectra+photometry without either component dominating the fit.

Four examples of the best-fitting model spectral energy distributions (SEDs) are shown in
Fig. \ref{tcm_fits_plot}, where the optical
galaxy spectra are plotted in black, the error spectra are plotted in
green, the photometric data-points are plotted in red and the best-fitting models are
plotted in grey. For each object, the left-hand panel of each row shows the full fit and the
right-hand panel shows a close-up of the fitting region containing the
optical spectra. For presentation purposes the optical spectra have
been smoothed using a 5-pixel boxcar.

\subsubsection{Adopted stellar masses}
In Fig. 3 we show the stellar-mass distributions for the K-Bright
sample, as derived using (left-to-right) $\tau-$models and DB
models based on the CB07 template libraries, followed by the
corresponding $\tau-$models and DB models based on the BC03 template
libraries. Although, as expected, the models based  on the CB07
templates produce lower stellar-mass estimates than the models based
on the BC03 templates (and the $\tau-$models produce lower
stellar-mass estimates than the DB models), perhaps the most striking
feature of Fig. 3 is the similarity between the resulting stellar-mass
distributions. This is highlighted by the fact that even between the
two extremes, the median stellar mass of the K-Bright sample only
changes by 0.17 dex. It is clear from Fig. 3 that the TP-AGB phase is
not having a significant impact on the derived stellar  masses of the
K-Bright sample. Furthermore, the similarity of the derived
stellar-mass distributions implies that the proceeding exploration  of
the size-mass relation at $z\simeq 1.4$ will not be dominated by
systematic stellar-mass uncertainties driven by galaxy SED template
choice.  Armed with this information, throughout the rest of the paper
we simply adopt the stellar-mass and age measurements derived from the BC03
$\tau-$models, in order that they can be directly compared to numerous
other studies in the literature (see Section 4.5).

\subsection{Galaxy size measurements}
Measurements of the galaxy radii were derived using two-dimensional
modelling of  the latest (DR8) $K-$band imaging of the UDS which has a
$5\sigma$ detection limit of $K=24.7$ (2\asec\,diameter apertures) and
0.75\asec\, FWHM image quality. Over the  redshift range covered by
the K-Bright sample ($1.3 \leq z \leq 1.5$) the UDS $K-$band imaging is centred on 
rest-frame wavelengths of $8800$\AA$~< \lambda < ~9600$\AA, which is
advantageous for two reasons.  Firstly, at these rest-frame
wavelengths the $K-$band imaging provides a size measurement
significantly long-ward of the 4000\AA\, break, where the galaxy SED
should be  relatively unaffected by any recent episodes of
star-formation. Secondly, at $z\simeq 1.4$ the UDS $K-$band imaging
provides size measurements well matched in terms of rest-frame
wavelength to the benchmark galaxy size measurements from Shen et
al. (2003), which are based on $z-$band SDSS imaging.

The galaxy radii were calculated using a modified version of {\sc
  galapagos} (Barden et al 2012) and {\sc galfit} \citep{peng}. {\sc
  galapagos} is a wrapper script for {\sc galfit} which fits single
S\'ersic profiles to each object in the image, taking initial
parameters from a catalogue generated by SExtractor (Bertin \& Arnouts 1996). A key advantage
of {\sc galapagos} compared to {\sc galfit} alone  is that it has an
algorithm for background estimation and, if necessary, will
simultaneously  model any nearby companion objects which would
otherwise unduly influence the fit.  

For this study the code was adapted to produce a densely sampled grid
of $R_e$, S\'ersic index and background level in order to estimate
accurate errors on the fitted parameters from contours of constant
$\Delta\chi^2$. It is worth emphasising that, because both $R_{e}$ and S\'ersic
index correlate with the sky-background level, including a variable 
background in the parameter estimation grid is essential for obtaining
realistic uncertainties on the derived parameters.

\begin{figure}
\includegraphics[scale=0.65,angle=270]{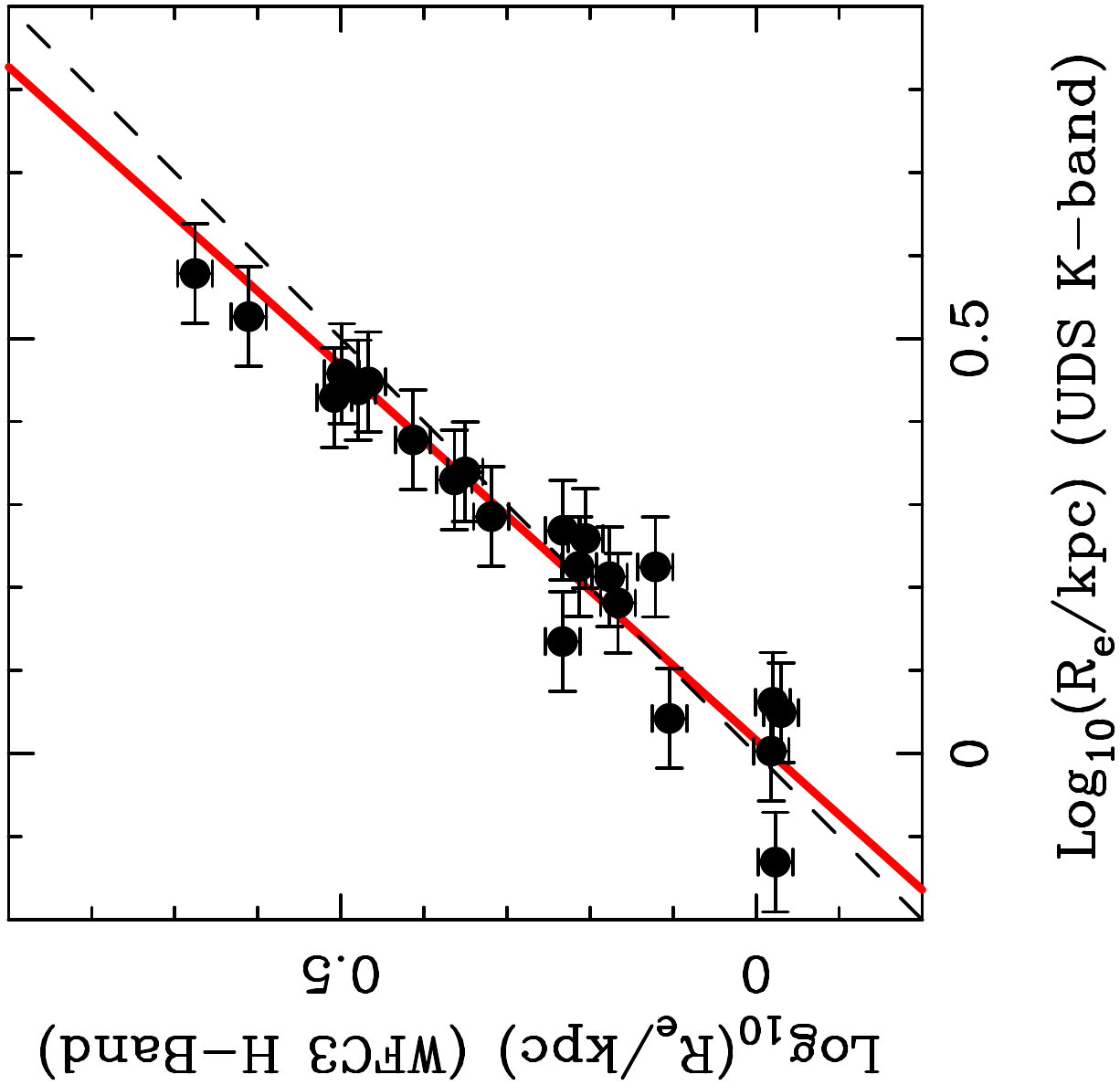}
\caption{A comparison between the effective half-light radii derived from ground-based $K-$band imaging and
HST $H-$band imaging for a sample of N=22 objects with $K_{tot}\leq 21.5$ and 
$1.3\leq z_{phot} \leq 1.5$ within the CANDELS/UDS field. The thick red 
line shows the best-fitting relationship (Equation 1) and the thin dashed line is 
a one-to-one relationship (see text for a full discussion).}
\end{figure}

\subsubsection{Comparison with higher resolution imaging}
One approach to testing the accuracy of our galaxy size measurements would be to analyse simulated images containing artificial
galaxies spanning a range of half-light radii and S\'{e}rsic index. We have not performed such a simulation in this work, 
simply because building and then reclaiming the properties of artificial galaxies with the 
same axi-symmetric galaxy models is likely to over-estimate the accuracy with which the galaxy sizes can be recovered. 

However, within the UDS field we are in the fortunate position that it is possible to directly compare galaxy sizes derived from
ground-based and HST imaging in the near-IR. Consequently, in order to test whether or not our determinations of the
galaxy radii have been affected by the ground-based spatial
resolution of the UDS $K-$band imaging, we have made use of the publicly available Cosmic
Assembly Near-infrared Deep Extragalactic Legacy Survey (CANDELS,
Grogin et al. 2011; Koekemoer et al. 2011) imaging in the UDS. The CANDELS dataset in the UDS features
$J-$ and $H-$band WFC3/IR imaging (FWHM $\simeq 0.2$\asec) over an
area  of $\simeq 200$ sq. arcmins to a $5\sigma-$depth of $26.8$
(0.6\asec diameter apertures).

Based on the photometric redshift catalogue of Bruce et al. (2012), there
are twenty-two galaxies within the CANDELS/UDS area with $K_{tot}\leq 21.5$ and photometric redshifts 
in the range $1.3\leq z \leq 1.5$. For this sample the half-light radii and S\'{e}rsic index 
measurements derived by Bruce et al. (2012), based on a {\sc galfit} analysis of 
the $H-$band HST imaging, were compared to those derived using the ground-based $K-$band imaging.

As can been seen from Fig. 4, the results of this comparison demonstrate that the two sets of half-light radii
measurements are well correlated ($r_{s}=0.95$) and follow an essentially one-to-one relation. The 
best-fitting relation between the two measurements of half-light radius was found to be:
\begin{equation}
R_{\rm WFC3}=(0.96\pm{0.05})R_{\rm UDS}^{1.11\pm0.08}
\end{equation}
where $R_{\rm WFC3}$ and $R_{\rm UDS}$ are the circularised half-light radii (in kpc) derived from the 
WFC3/IR and ground-based data respectively. Moreover, the corresponding S\'{e}rsic index measurements were 
also found to be well correlated ($r_{s}=0.82$), with only 1/22 objects returning a significantly different 
best-fitting S\'{e}rsic index ($n_{\rm{WFC}}=2.0$ versus $n_{\rm{UDS}}=3.5$). Consequently, for the purposes of the proceeding analysis the
measurements of half-light radii and S\'{e}rsic indices derived from the ground-based $K-$band imaging 
are adopted without correction.

\section{The Stellar Mass-Size Relation}
\begin{figure*}
\begin{center}
\includegraphics[scale=0.7]{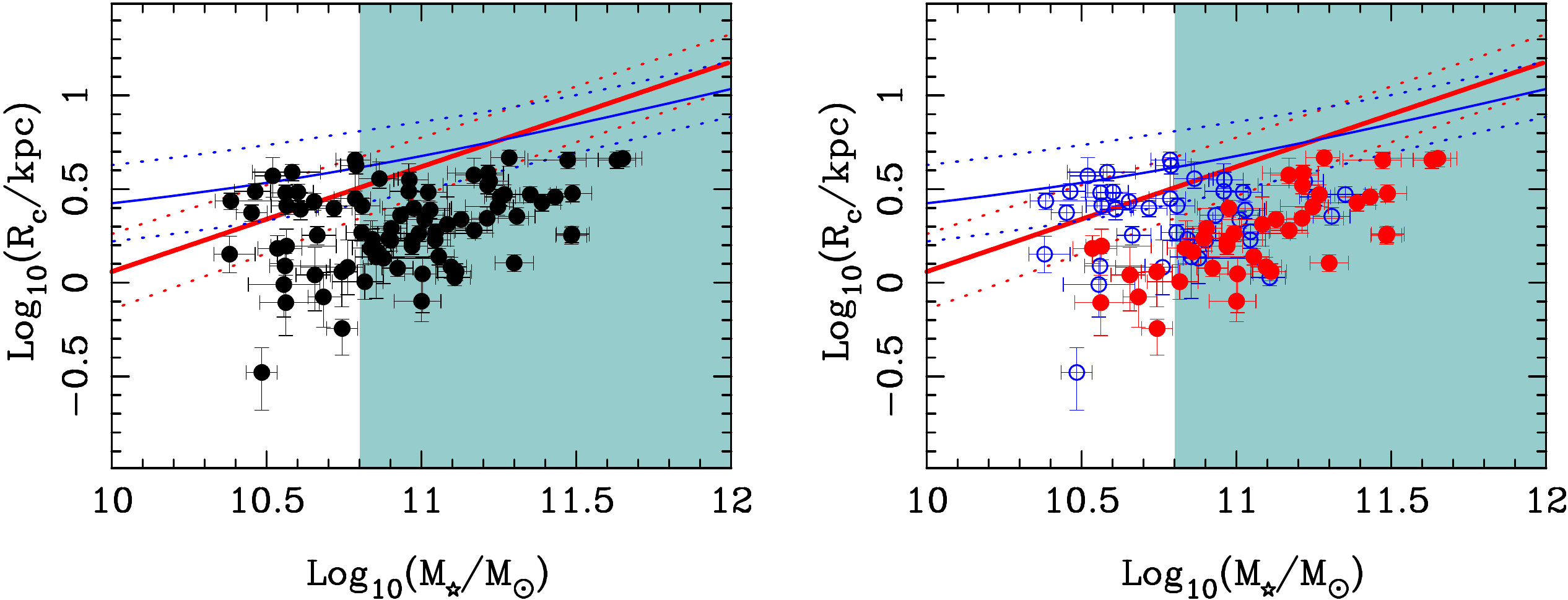}
\caption{The galaxy size - stellar mass relation for the K-Bright
  sample. The left-hand panel shows the location of the K-Bright
  sample on the half-light radius - stellar mass plane. The thin
  (blue) and thick (red) solid lines show the  relations derived for
  $z<0.1$ SDSS galaxies by Shen et al. (2003) for late-type ($n<2.5$)
  and early-type ($n\geq2.5$)  galaxies respectively (dotted lines
  indicate the associated $1\sigma$ scatter). 
The shaded region at $\log(\rm{M_{\star}}/\Msun)\geq10.8$ highlights where the K-Bright sample should be 100\% mass-complete, even for the most passive galaxies
(see text for discussion). The right-hand panel
  shows the same information, except that the K-Bright galaxies have
  been split into passive (red solid circles) and  star-forming (blue
  open circles) sub-samples depending on whether their specific
  star-formation rate (sSFR) lies  above or below 0.1 Gyr$^{-1}$. In both panels, and all subsequent figures, the
  plotted half-light radii measurements have been circularized
  (i.e. $R_{c}=\sqrt{(b/a)}R_{e}$) in order to facilitate direct
  comparison with the results of Shen et al. (2003).}
\end{center}
\end{figure*}

In this section we present the basic results of our study of the
galaxy size-mass relation at $z\simeq 1.4$, exploring the location of the K-Bright sample on
the size-mass plane as a function of specific star-formation rate,
morphology and stellar population age. In what follows we will
consistently compare the results for the K-Bright sample 
with the size-mass relations derived by Shen et al. (2003)  for
early ($n\geq 2.5$) and late-type ($n<2.5$) galaxies in the SDSS. In
order to perform this comparison we have converted the  semi-major
axis half-light radii ($R_{e}$) derived by {\sc galfit} to their
equivalent circularized values ($R_{c}$) using the  standard
conversion: $R_{c}=\sqrt{(b/a)}R_{e}$, where $a/b$ is the best-fitting
axial ratio.

In the left-hand panel of Fig. 5 we show the full K-Bright sample on
the size-mass plane. Also shown are the size-mass relations  for
late-type (thin blue line) and early-type (thick red line) galaxies
in the SDSS as derived by Shen et al. (2003), with the  dotted lines
indicating the $1\sigma$ scatter associated with both relations. The
shaded region at $M_{\star}\geq 6\times 10^{10}\Msun$ highlights the r\'{e}gime
where the K-Bright sample is fully mass
complete. This mass completeness threshold has been calculated in a
deliberately conservative manner and is based on a model galaxy at
$z=1.4$ with $K_{tot}=21.5$, zero current star formation and an age of
4 Gyr.  Consequently, at $M_{\star}\geq 6\times 10^{10}\Msun$
the K-Bright sample should be mass complete, even for entirely passive
galaxies. In the stellar mass interval $3\times 10^{10}\Msun \leq M_{\star} \leq 6\times 10^{10}\Msun$ the K-bright sample may be partially incomplete
for the most passive galaxies, but should still be complete for
actively star-forming galaxies, unless they are heavily reddened. The
apparent cut-off at stellar masses of $M_{\star}\simeq 3 \times 10^{10}\Msun$
is thus a function  of the adopted $K_{tot}=21.5$ magnitude limit and the
red colour selection of the FORS2 spectroscopic sample (i.e. $i-K\geq 2$).

\subsection{Specific star-formation rate}
In the right-hand panel of Fig. 5 the K-Bright sample has been split on the basis 
of specific star-formation rate (sSFR), where objects with
sSFR $\leq 0.1$ Gyr$^{-1}$ are plotted as red filled circles and
objects with sSFR $> 0.1$ Gyr$^{-1}$ are plotted as open blue
circles. Although it is obviously possible to derive a value of sSFR
directly from the best-fitting SED templates, in order to avoid too much coupling between
the adopted stellar mass and star-formation rate measurements, the decision was taken to exploit the 
different empirical star-formation indicators available for the K-Bright sample. 

The first star-formation indicator we have employed is the UV luminosity calibration of 
Madau et al. (1998) and was based on the observed $B-$band magnitudes, dust corrected using the best-fitting value of $A_{\rm v}$.
The second indicator is based on the [OII] emission line fluxes (or upper limits), and is based on the Kennicutt (1998) calibration assuming that the H$\alpha$ emission line undergoes an average extinction of one magnitude. 
The final star-formation indicator is based on the observed $24\mu$m fluxes (or upper limits), which are initially 
converted into rest-frame $8\mu$m luminosities using a $10^{11}L_{\odot}$ starburst template from Lagache et al. (2003). 
Based on the $8\mu$m luminosity the total infra-red luminosity is then estimated using the calibration of 
Bavouzet et al. (2008) and the star-formation rate is estimated using the calibration of Kennicutt (1998).

The adopted star-formation rate estimate for each 
member of the K-Bright sample is taken as the {\it largest} value 
derived from [OII] emission-line flux, dust-corrected UV luminosity and MIPS 24$\mu$m flux. 
In this fashion we have deliberately adopted the {\it maximum} likely value of
sSFR for each object, which should ensure that objects with extremely
low values of sSFR can be regarded as genuinely passive. The choice
of $0.1$ Gyr$^{-1}$ as the threshold between actively star-forming and
passive objects is motivated by several different
factors. Fundamentally, this value is the median sSFR value of the
K-Bright sample and is also a suitable dividing line
separating galaxies which are on or off the so-called ``main sequence
of star-formation'' at this epoch (e.g. Elbaz et al. 2011). Moreover, at $z\simeq 1.4$
a sSFR value of $0.1$ Gyr$^{-1}$ implies that a galaxy will require a
Hubble time in order to double its stellar mass.

\begin{figure*}
\begin{center}
\includegraphics[scale=0.7]{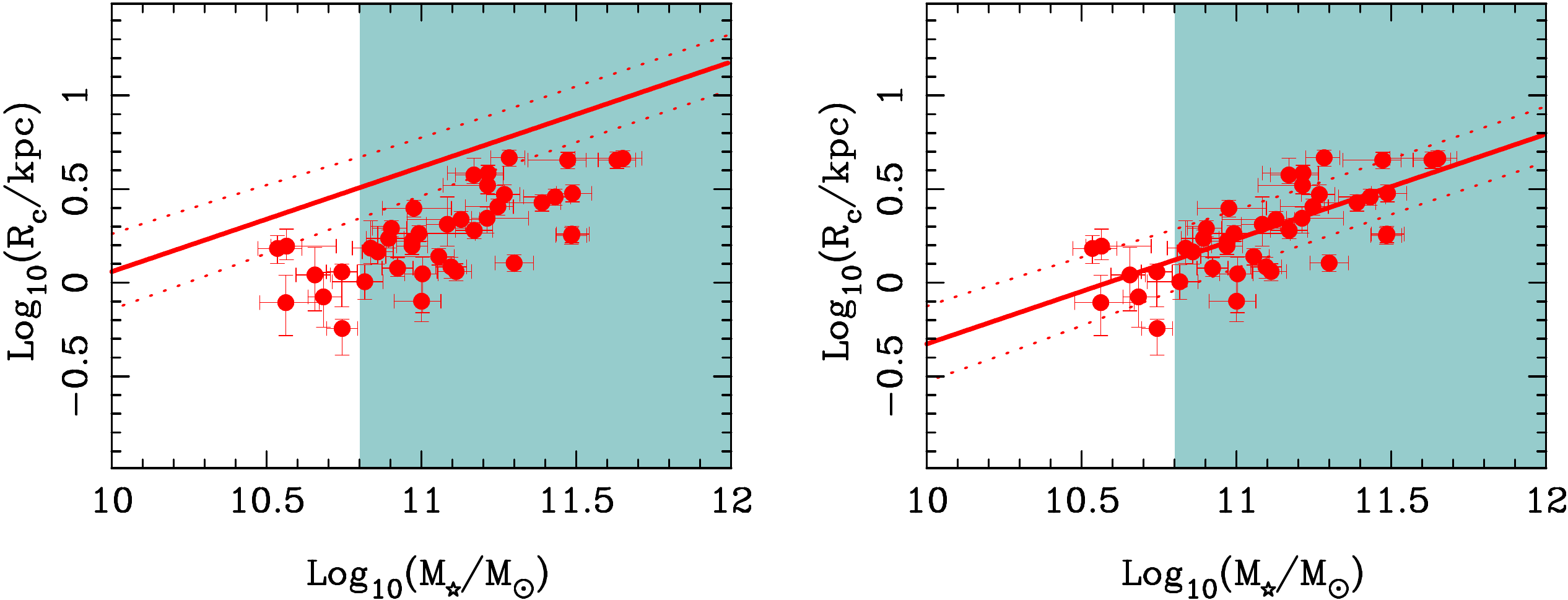}
\caption{The left-hand panel shows the galaxy size - stellar mass
  relation for those members of the K-Bright sample classified as
  passive according to our adopted sSFR $\leq 0.1$ Gyr$^{-1}$
  threshold. The solid and dotted lines show the size-mass relation
  for early-type galaxies in the SDSS derived by Shen et al. (2003)
  and the corresponding $1\sigma$ scatter respectively. It can be seen
  that the passive K-Bright galaxies follow a
  size-mass relation which has an identical slope (and scatter) to that of local early-type galaxies, but
 is simply shifted in normalisation. This is illustrated by the right-hand panel which
  shows the effect of reducing the vertical normalization of the Shen
  et al. early-type relation by a growth factor of
  $f_{g}=2.43\pm{0.20}$ (see text for discussion).}
\end{center}
\end{figure*}

\begin{figure*}
\begin{center}
\includegraphics[scale=0.7]{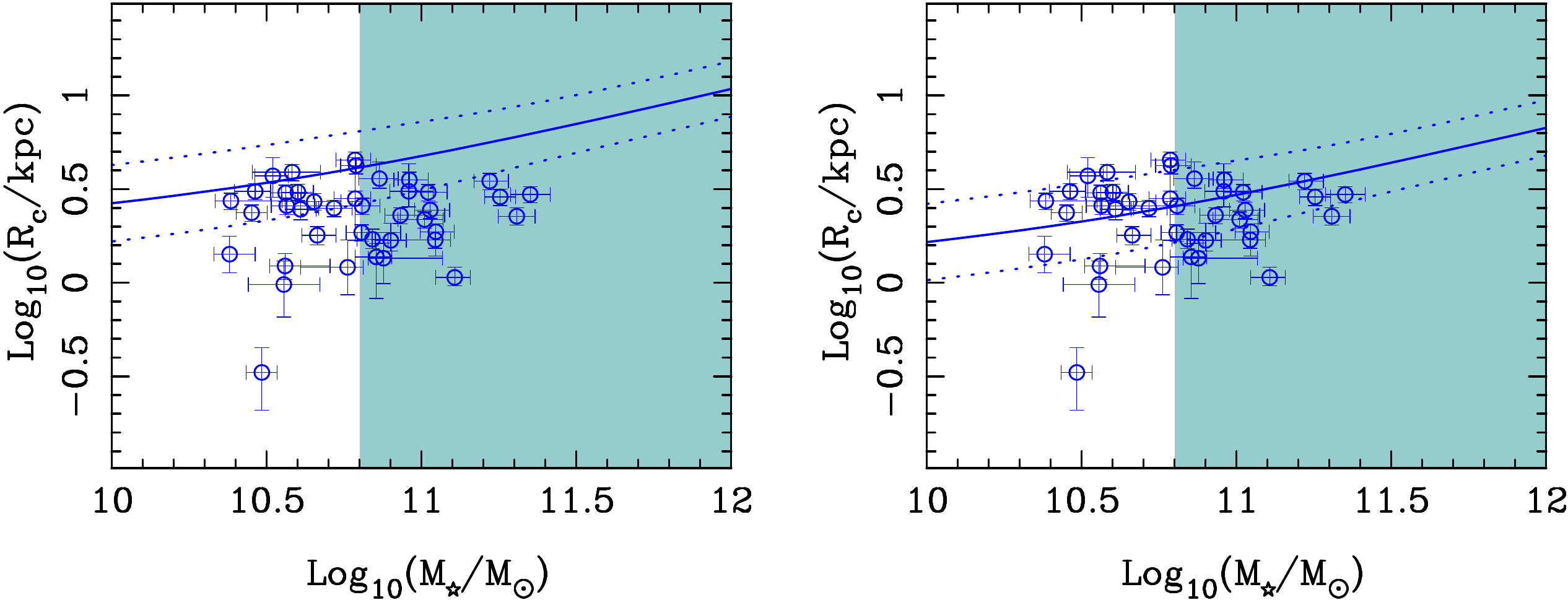}
\caption{The left-hand panel shows the galaxy size - stellar mass
  relation for the those members of the K-Bright sample classified as
  star-forming according to our adopted sSFR $> 0.1$ Gyr$^{-1}$
  threshold. The solid and dotted lines show the size-mass relation
  for late-type galaxies in the SDSS derived by Shen et al. (2003) and
  the corresponding $1\sigma$ scatter respectively. It can be seen
  that the star-forming K-Bright galaxies are consistent with
  following a size-mass relation similar to that of local late-types, but shifted in
  normalisation. This is illustrated by the right-hand panel which
  shows the effect of reducing the vertical normalization of the Shen
  et al. late-type relation by a growth factor of
  $f_{g}=1.61\pm{0.17}$ (see text for discussion).}
\end{center}
\end{figure*}

\subsubsection{Passive galaxies}
It can clearly be seen from the right-hand panel of Fig. 5 that the
passive members of the K-Bright sample follow a distinct
size-mass relation from the actively star-forming members. This point
is clarified further in Fig. 6 which shows just the passive members of the 
K-Bright sample on the size-mass plane. It is apparent from the
left-hand panel of Fig. 6 that the passive members of the K-Bright
sample follow a well defined size-mass relation, with a slope
consistent with the size-mass relation displayed by local early-type
galaxies, but with systematically lower half-light radii than their
local counterparts. 

In this context it is instructive to calculate the size growth factor ($f_{g}$),
defined here as the increase in half-light radius required to place the
K-Bright galaxies onto the appropriate local galaxy size-mass relation from 
Shen et al. (2003). To reconcile the passive members of the K-bright sample with the local
early-type size-mass relation requires a median growth factor of
$f_{g}=2.43\pm 0.20$, even allowing for no increase in stellar mass. This off-set is
illustrated by the right-hand panel of Fig. 6, where the local
size-mass relation for early-type galaxies derived  by Shen et
al. (2003) has been lowered in normalization by a factor of
$f_{g}=2.43$. Remarkably, it can be seen that this simple
renormalization of the local early-type size-mass relation provides an
excellent description of the size-mass relation followed by the passive members of the K-bright sample
at $z\simeq 1.4$, both in terms of slope and scatter. We will return
to this point in Section~5. 

\begin{figure}
\begin{center}
\includegraphics[scale=0.65]{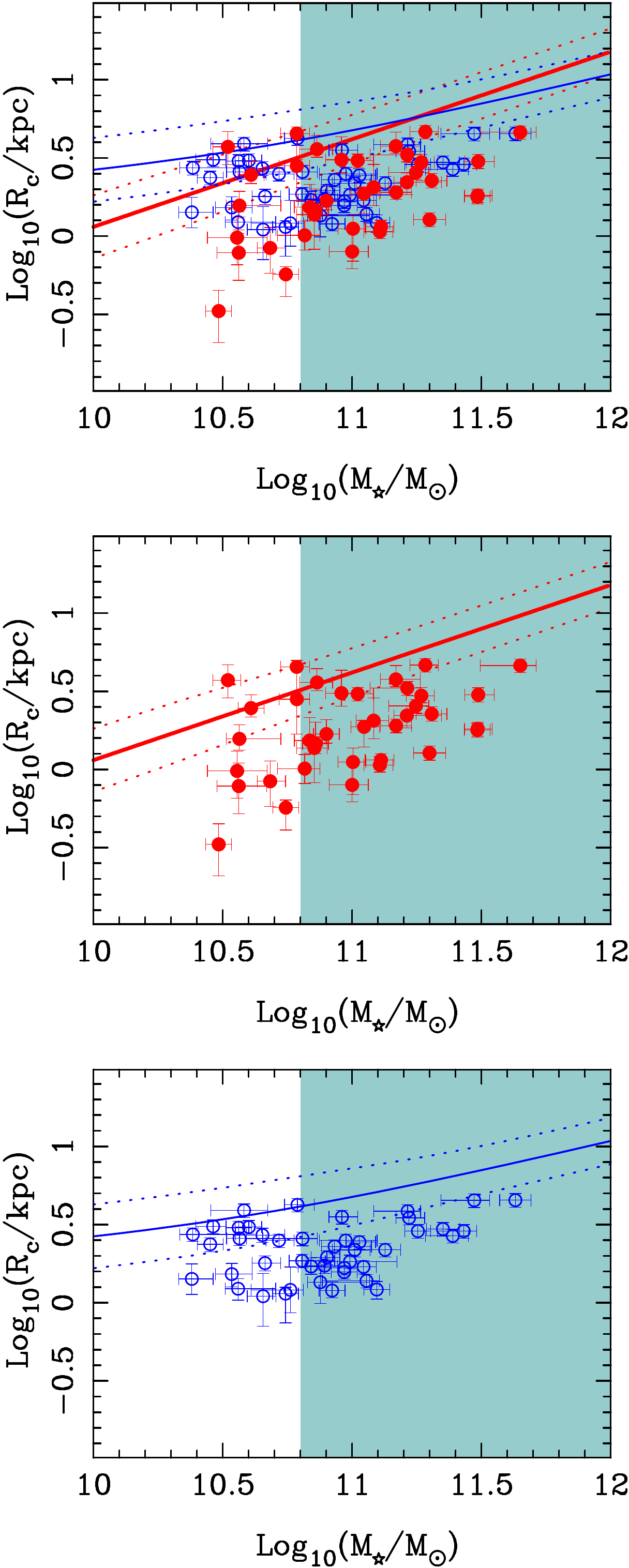}
\caption{The top panel shows the size-mass relation for the
  K-Bright sample with early-type galaxies ($n\geq 2.5$) plotted as solid red circles and 
late-type galaxies ($n<2.5$) plotted as open blue circles. The middle panel shows the early-type K-Bright 
galaxies compared to the local early-type size-mass relation from Shen et al. (2003). The bottom panel 
shows the equivalent comparison for the late-type K-Bright galaxies.}
\end{center}
\end{figure}

\subsubsection{Star-forming galaxies}
In the left-hand panel of Fig. 7 we plot the star-forming (sSFR $>
0.1$ Gyr$^{-1}$) half of the K-Bright sample on the size-mass plane
and compare with the local size-mass relation for late-type galaxies
in the SDSS derived by Shen et al. (2003). In contrast to the passive members of the
K-Bright sample, it can immediately be seen that $\simeq 50\%$ of the
star-forming galaxies are already consistent with the size-mass relation of local late-type galaxies.
As a result, the median value of the growth factor for the star-forming members of the 
K-Bright sample is only $f_{g}=1.61\pm0.17$ \footnote{The outlier at ($\log(\rm{M}/\Msun)=10.5, \log(R_{c}/kpc)=-0.5$) has been excluded from this calculation.}, as illustrated in the right-hand panel of Fig. 7. 

The results of this sub-section are broadly consistent with many
previous studies in the literature (see Section 1) in that, at a given epoch,  the
off-set from the local size-mass relations is significantly larger for
passive galaxies than for their star-forming contemporaries. 
A detailed comparison with relevant results from recent literature will be presented in Section 4.5.

\subsection{Morphology}
When considering the evolution of the galaxy size-mass relation it is
conventional to consider the influence of galaxy morphology. Indeed,
much of the discussion in the literature has concentrated on the size
evolution of morphologically  selected early-type
galaxies. Consequently, in this section we investigate the size-mass
relation for the K-Bright sample, splitting the sample into late-type
and early-type sub-samples using the same S\'{e}rsic index threshold
of $n=2.5$ adopted by Shen et al. (2003).

\subsubsection{Early-type galaxies}
The middle panel of Fig. 8 shows the size-mass relation for the early-type ($n\geq
2.5$) members of the K-Bright sample compared to the size-mass
relation for local early-type galaxies from Shen et al. (2003).  It
can be seen that the early-type galaxies follow a size-mass relation
which has a slope which is  consistent with the local size-mass
relation for early-types, but is off-set to smaller radii at a  given
stellar mass. In terms of the growth factor ($f_{g}$), the early-type
galaxies are off-set  from the local size-mass relation by a factor of
$f_{g}=2.37\pm{0.29}$.

If we confine our attention to the growth factor required to reconcile
them with the local size-mass relation for early-type galaxies,  it is
clear that the passive and early-type sub-samples of the K-Bright
galaxies are very similar, with $f_{g}=2.43\pm{0.20}$ and
$f_{g}=2.37\pm{0.29}$ respectively. However, it is noticeable that the
scatter associated with the size-mass relation  displayed by the
early-type sub-sample is somewhat larger than that associated with the
size-mass relation of the  passive galaxy sub-sample. Comparing Fig. 6
with Fig. 8 it is clear that the reason for the increased scatter is
that, unlike the passive galaxy sub-sample, $\simeq 30\%$ of the early-type 
sub-sample have half-light radii which are consistent with the 
local early-type size-mass relation. 

Interestingly, although splitting the full K-Bright sample at $n\geq
2.5$ or sSFR $ \leq0.1$ Gyr$^{-1}$ delivers  similar sized sub-samples
(N=37 and N=41 galaxies respectively), the correspondence between an early-type
morphology  and passivity is not one-to-one. Indeed, the forty-one galaxies
in the K-Bright sample classified as passive  are split relatively
evenly between early-type (N=23) and late-type morphologies
(N=18). Similarly, of the thirty-seven objects in the early-type sub-sample,
only twenty-three are also classified as passive. Consequently, it is clear
that a relatively large fraction ($\simeq 20\%$) of the K-Bright
galaxies are classified as star-forming early-types according to our
adopted sSFR and S\'{e}rsic index criteria.

\subsubsection{Late-type galaxies}
In the bottom panel of Fig. 8 we plot the size-mass relation for the 
late-type ($n<2.5$) members of the K-Bright sample. As with the early-type sub-sample, the 
late-type members of the K-Bright sample appear to follow a size-mass relation which has a slope 
which is consistent with the equivalent local size-mass relation, but off-set to smaller half-light 
radii at a given stellar mass. If we calculate the off-set from the local size-mass relation in terms of
the growth factor we derive a median value of $f_{g}=2.15\pm 0.15$.

It is noteworthy that this growth factor is larger than the growth 
factor previously calculated for the star-forming members of the K-Bright 
sample ($f_{g}=1.61\pm0.17$). Again, the reason for this apparent discrepancy is that the correspondence 
between late-type morphology and active star-formation is not one-to-one. In fact, of the forty-four members of 
the late-type sub-sample, twenty-six are actively star-forming galaxies and eighteen are galaxies which are 
classified as passive according to our adopted sSFR threshold. The overall growth factor of 
$f_{g}=2.15\pm 0.15$ is therefore effectively an average of the $f_{g}=1.79\pm{0.17}$ displayed 
by the star-forming late-type galaxies and the significantly larger growth factor of $f_{g}=2.48\pm{0.23}$ 
displayed by the {\it passive} late-type galaxies.

Therefore, not only is $\simeq 20\%$ of the K-Bright sample comprised
of star-forming bulges which are  {\it larger} than their passive
counterparts, a further $\simeq 20\%$ is comprised of passive disk-like galaxies 
which are {\it smaller} than their actively star-forming
counterparts. In combination with the  tight size-mass correlation
displayed by the passive sub-sample in Fig. 5, this information leads
naturally to the conclusion that the location of massive $z\simeq 1.4$
galaxies on the size-mass plane is a stronger function of specific
star-formation rate than galaxy morphology.

Finally, it is interesting to note that $\geq 40\%$ of the passive members of the K-Bright 
sample are classified as having a disk-like morphology (i.e. $n<2.5$). This result provides potentially 
important information about the quenching of star-formation in massive galaxies and will be discussed in more detail 
in Section 5.

\subsection{Stellar population age}
In this section we examine the evidence for a relationship between
galaxy size and stellar  population age within the early-type/passive
members of the K-bright sample, given that the  confirmation of any
such relationship would offer important constraints on competing
evolutionary  scenarios. Recent results on this issue in the
literature have been controversial, with different studies  finding
apparently contradictory claims (e.g. Saracco et al. 2009; Trujillo et
al. 2011).  In the following discussion we have adopted the
best-fitting age since formation ($t_{for}$) as our age indicator,
although quantitatively similar results are derived if the alternative
age indicator $t_{for}/\tau$ is employed.

\subsubsection{Passive galaxies}
The median age of the passive galaxy sub-sample is $3.3\pm{0.2}$ Gyr, 
which is consistent with a formation redshift of
$z_{for}\simeq 4.5$. If we divide the sample into ``old'' and
``young'' based on this median age, then the passive sub-sample also
cleanly separates in terms of half-light radius and stellar mass. For
example, the median stellar mass of the ``old'' passive galaxies is
$M_{\star}=(1.64\pm{0.29})\times 10^{11}\Msun$, whereas the median stellar mass of
the ``young'' passive galaxies is $M_{\star}=(0.79\pm{0.21})\times 10^{11}\Msun$.
The corresponding median values for the half-light radii are
$2.52\pm{0.35}$ kpc for the ``old'' passive sub-sample and
$1.55\pm{0.22}$ kpc for the ``young'' passive sub-sample.

Several aspects of these results are noteworthy. Firstly, the fact that the
oldest passive members of the K-Bright sample are systematically  more
massive than the youngest passive members indicates that the process
of ``downsizing'' is already in place for massive galaxies  at
$z\simeq 1.4$. Moreover, the difference in median
half-light radius for the ``old'' and ``young'' passive  galaxies is
entirely consistent with the correlation between half-light radius and
stellar mass seen for early-type galaxies in the  SDSS by Shen et
al. (2003), i.e. $r_{hl}\propto M^{0.56}$ (as expected given Fig. 6).

Therefore, although we do find a factor of $\simeq 1.6$ difference in
the median half-light radii of  the ``old'' and ``young'' passive
galaxy sub-samples, this is simply a consequence of the correlations
between size-mass and  mass-age. Indeed, because the passive galaxy
sample follows a size-mass relation with an identical slope to that seen at
low-redshift  (see Fig. 6), the median growth factors of the ``old''
and ``young'' members of the passive galaxy sub-sample are perfectly
consistent ($f_{g}=2.39\pm{0.32}$ and $f_{g}=2.41\pm{0.26}$
respectively). Based on this evidence we conclude that there is no
correlation between stellar population age and off-set from the local size-mass 
relation for passive galaxies within the K-Bright sample (see Section 5 for further discussion).

\subsubsection{Early-type galaxies}
A more direct comparison with existing studies in the literature is to
examine if there are any correlations with stellar population age
within the early-type ($n\geq 2.5$) members of the K-Bright sample.

To explore this issue, we adopt the same simple strategy and split the
early-types into ``young'' and ``old'' sub-samples using the median
age since formation (which is again 3.3 Gyr).  If we then
calculate the median growth factor needed to reconcile the K-Bright
early-types with the local early-type size-mass  relation we derive
values of $f_{g}=2.50\pm{0.32}$ for the ``old'' sub-sample and
$f_{g}=2.28\pm{0.46}$ for the ``young'' sub-sample. As for the
passive members of the K-Bright sample, based on these results we
conclude that there is no indication that the off-set from the local
early-type size-mass relation is related to stellar population age at
$z\simeq 1.4$ (consistent with the conclusions of Cimatti, Nipoti \& Cassata 2012). 
In fact, as with the passive sub-sample, the median half-light radius of the  ``old''
early-type galaxies is actually larger than that of the ``young'' early-types
($2.2\pm0.3$ kpc versus $1.6\pm{0.3}$ kpc), as a result of the correlation between stellar population age and stellar
mass.

\subsection{Dynamical mass measurements}

\begin{figure}
\begin{center}
\includegraphics[scale=0.7]{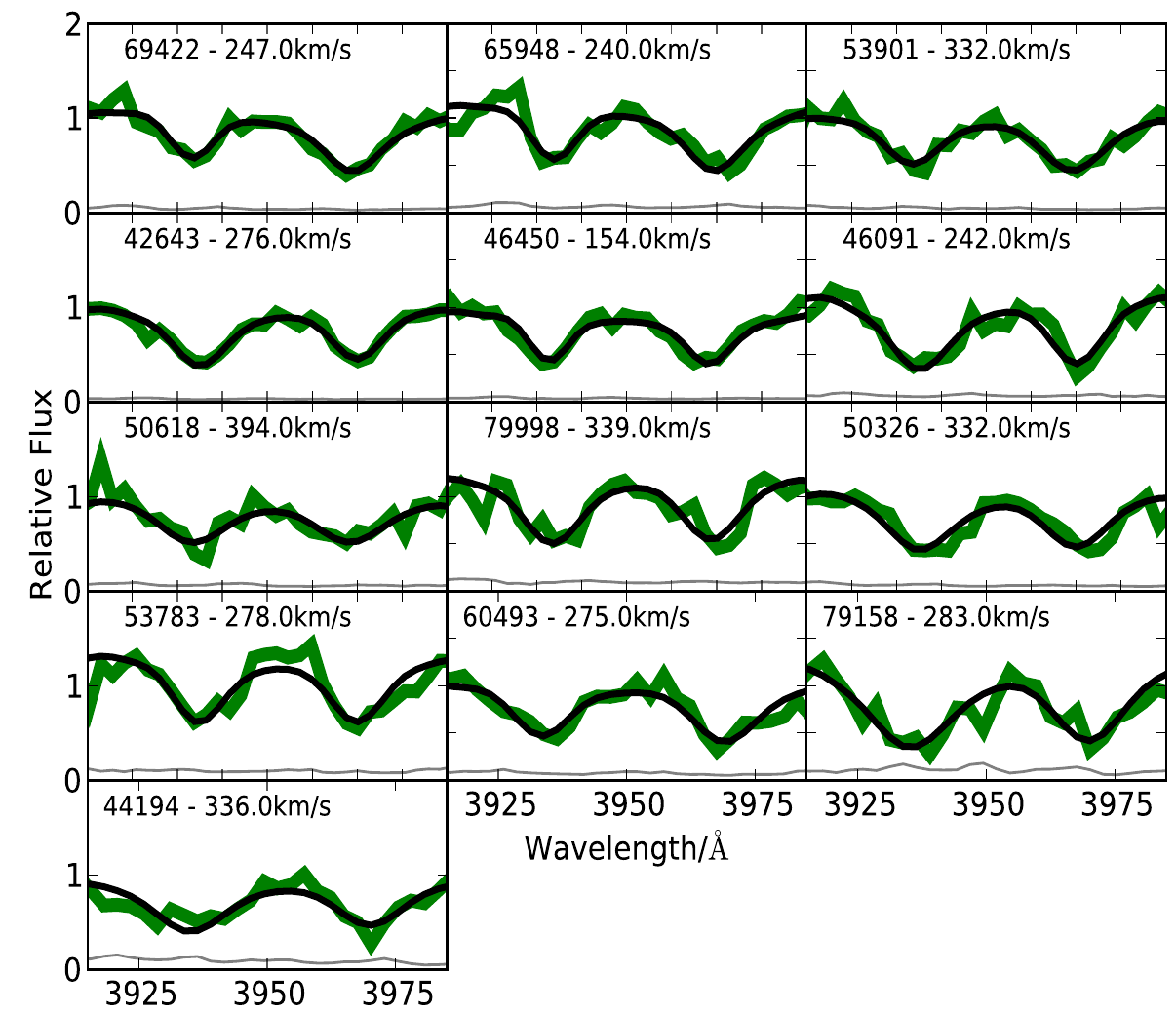}
\caption{Illustration of the model fitting employed to derive stellar-velocity 
disperions for thirteen galaxies with sufficiently high signal-to-noise (SNR $\simeq 8$ per pixel) in
the Ca H \& K spectral region. In each panel the thick green line is the galaxy spectrum, the black line
is the best-fitting convolved stellar spectrum and the thin grey line is the error spectrum. The catalogue 
ID number and best-fitting stellar-velocity dispersion are shown at the top of each panel.}
\end{center}
\end{figure}

\begin{figure}
\begin{center}
\includegraphics[scale=0.7]{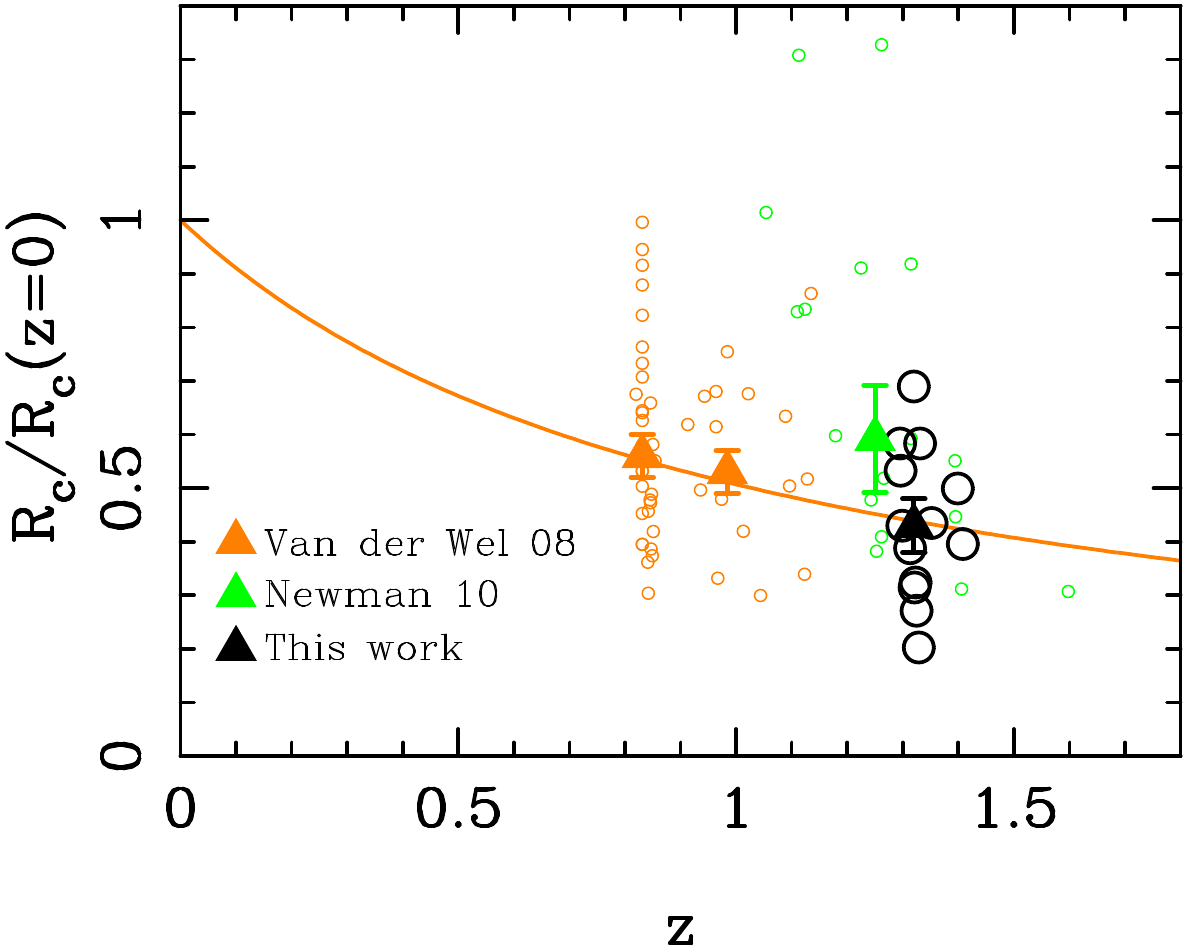}
\caption{The evolution of circularised effective radius as a function
  of redshift for early-type galaxies with dynamical mass estimates
  of $M_{dyn}\simeq 10^{11}\Msun$, normalized to the dynamical mass -
  size relation determined from the SDSS by van der Wel et al. (2008);
  see  text for details. The orange data points are taken from the
  sample of 50 early-type galaxies published by van der Wel et
  al. (2008), which contains both field galaxies at $z\simeq 1$ and
  cluster galaxies at $z=0.831$. The green data points are a sample of
  17 early-type field  galaxies in the redshift interval
  $1.054 \leq z \leq 1.598$ with velocity dispersion measurements derived from
  Keck spectroscopy published by Newman et al. (2010). The black data
  points are the 13 galaxies from the K-Bright sample for which it was
  possible to obtain velocity dispersion measurements and therefore
  dynamical mass estimates. In each case the filled triangles  with
  error bars indicate the sample medians. The solid curve shows the
  evolutionary trend derived by van der Wel et al. (2008) which has
  the  functional form $R_{c}(z) \propto (1+z)^{-0.98\pm0.11}$. In
  order to agree with the velocity dispersion normalization adopted
  both here and in van der Wel et al. (2008), the velocity dispersion
  measurements of Newman et al. (2010) have been corrected to match
  the velocity dispersion  within the half-light radius according to
  the prescription outlined in Jorgensen et al. (1996).}
\end{center}
\end{figure}
For a thirteen object sub-sample of the K-Bright galaxies it was possible to
extract a reliable measurement of their stellar-velocity dispersion. Consequently, 
for this small sub-sample it is possible to derive dynamical-mass estimates and thereby provide
an independent measurement of the typical size evolution, free from the systematic 
uncertainties associated with stellar-mass measurements based on SED fitting. The sub-sample of
galaxies with stellar-velocity dispersion measurements is
confined to those objects which were bright enough to provide the
necessary signal-to-noise (i.e. SNR $\simeq 8$ per pixel) and to the redshift 
interval $1.30 \leq z \leq 1.41$, such that the strong Ca H \&
K absorption features lie within a clean region of spectrum, clear of
strong sky-line residuals. 

The velocity dispersions were measured via $\chi^{2}$ minimization in
pixel space, using a template library of 215 high signal-to-noise
stellar spectra taken from STELIB Stellar Library (Le Borgne et
al. 2003), as illustrated in Fig. 9.  The corresponding dynamical mass estimates were derived
using the following expression:
\begin{equation}
M_{dyn} = \frac{5\sigma^2R_c}{G}
\label{dm}
\end{equation}
\noindent
where $R_{c}$ is the circularised half-light radius and $\sigma$ is the line-of-sight velocity dispersion within the
half-light radius (aperture corrected according to Jorgensen et al. 1996).

\subsubsection{Size evolution}
Based on the relationship between half-light radius and dynamical mass
derived for early-type SDSS galaxies by  van der Wel et al. (2008), it
is possible to predict the $z=0$ half-light radius of a galaxy with a
given dynamical mass.  Therefore, for each K-Bright galaxy with a
measured velocity dispersion it is possible to calculate the ratio of
the observed half-light radius at $z\simeq 1.4$ to the expected
half-light radius at $z=0$ for a galaxy of the same dynamical mass. 

In Fig. 10 we show the results of this calculation for the thirteen
K-Bright galaxies with velocity-dispersion measurements. In Fig. 10 we
also show the van der Wel et al. (2008) sample of 50 early-type
galaxies, consisting of both $z\simeq 1$ field galaxies and cluster
galaxies at $z=0.831$, and the  results from Newman et al. (2010) for
17 early-type galaxies in the redshift interval $1.054 \leq z \leq 1.598$. The
solid curve shown in Fig. 10 is the best-fitting relation for the
observed half-light radius evolution from van der Wel et al. (2008)
which has the form: $R_{c}(z) \propto (1+z)^{-0.98\pm0.11}$.

It can be seen from Fig. 10 that the results derived here are in good
agreement with those derived by  Van der Wel et al. (2008) and Newman
et al. (2010). Moreover, using the median value of $R_{c}/R_{c(z=0)}$ for the K-bright
sub-sample with dynamical-mass measurements we can derive an
independent estimate of the growth factor needed to reconcile the
sizes of the K-Bright galaxies with their local counterparts. It is
encouraging that the resulting value of $f_{g}=2.33\pm{0.32}$ is fully
consistent with our measurements derived on the basis of the
stellar mass - size relation.

\subsection{Comparison with the literature}
Given the extensive body of work dealing with the size evolution of
massive galaxies already in the literature, it is  clearly impossible
to comprehensively compare our results with every relevant paper on
the subject.  Consequently, in this sub-section we have chosen to
compare our results with four representative papers which cover the same 
redshift and stellar-mass range as the K-Bright sample.

\subsubsection{Trujillo et al. (2007)}
Trujillo et al. (2007) analysed the sizes of massive ($M_{\star}\geq
10^{11}\Msun$) galaxies in the  redshift interval $0.2<z<2.0$, based on
a sample of 810 $K-$band selected galaxies in the Palomar/DEEP-2 survey.
Using stellar-mass estimates based on the BC03
stellar population models (Chabrier IMF) and circularized  half-light radii derived
from the available F814W ACS imaging, Trujillo et al. observed rapid
size evolution in both the massive late-type ($n<2.5$) and early-type ($n\geq2.5$)
galaxy populations. Given that the Trujillo et al. sample is $K-$band
selected, adopts the same SED fitting models and covers the same
stellar-mass range as the K-Bright sample, it is clearly of interest
to compare their size evolution results with those derived here.

A precise comparison is slightly problematic because the redshift
range spanned by the K-Bright sample ($1.3 \leq z \leq 1.5$) straddles two of
the redshift bins adopted by Trujillo et al. ($1.1<z<1.4$ and
$1.4<z<1.7$). However, if we simply adopt the average of the results for the two 
Trujillo et al. redshift bins, their results suggest that at $z\simeq
1.4$ the growth factor for late-type galaxies is  $f_{g}\simeq 1.8$
and the growth factor for early-type galaxies is $f_{g}\simeq
3.3$. Exactly the same growth factors are obtained if we instead 
adopt the fitting formulas derived by Buitrago et al. (2008), who
incorporated the Trujillo et al. (2007) results  in their study of the
size evolution of massive galaxies over the redshift interval
$0.0<z<3.0$. These figures are to be compared with  the results
presented in Section 4.2, which were $f_{g}=2.15\pm{0.15}$ for late-types
and $f_{g}=2.37\pm{0.29}$ for early-types.

It is clear from this comparison that while our results for massive
late-type galaxies are in reasonable agreement, our determination of
the size evolution for massive early-type galaxies at $z\simeq 1.4$ is
significantly smaller than derived by Trujillo et al. (2007). 
However, in reality, the locus of the K-Bright early-type galaxies on the size-mass plane is 
entirely consistent with the Trujillo et al. sample (e.g. their Fig. 7).
The fundamental reason for the difference in derived growth factors is simply that the
K-bright sample does not display the significant ($\simeq 20\%$) tail of 
objects with $R_{c}\leq 1$ kpc which is present in the Trujillo et al. (2007) sample.
It is possible that the lack of $R_{c}\leq 1$ kpc objects in the K-Bright sample is associated 
with the limitations of ground-based imaging, although our comparison to the available WFC3/IR imaging 
(see Section 3.2.1) suggests that the $K-$band $R_{c}$ determinations are reliable down to $\simeq 1$ kpc and there is no
indication of a ``plateau'' in the $R_{c}$ values shown in Fig. 6.

\subsubsection{Saracco et al. (2009)}
Saracco, Longhetti \& Andreon (2009) investigated the massive galaxy
size-mass relation using $H-$band HST NICMOS imaging of a heterogeneous sample of 32, morphologically
classified, early-type galaxies in the redshift interval $1<z<2$.
Based on the available  $0.4\mu$m-2.2$\mu$m photometry, 
Saracco et al. (2009) performed SED fitting using the BC03 template
library (Chabrier IMF) and exponentially decaying SFHs.
Consequently, it is straightforward to directly compare the results of
Saracco et al. (2009) with those derived here.

Based on their SED fitting, Saracco et al. found that their
sample of early-type galaxies (ETGs) displayed a bi-model distribution
of ages, with peaks at $\simeq 1$ Gyr and $\simeq 3.5$ Gyr
respectively.  Splitting their sample into old ETGs (oETGs) and young
ETGs (yETGs) using an age threshold of 2 Gyr, Saracco et al.  found
that while $z\simeq 1.5$ oETGs have half-light radii which are a
factor of $\simeq 2.5$ smaller than their local counterparts, yETGs
at $z\simeq 1.5$ are fully consistent with the local size-mass
relation.  As already discussed in Section 4.3, amongst the early-type
($n\geq 2.5$) members of the K-Bright sample we find no evidence that
the off-set from the local size-mass relation is a function of stellar
population age, instead finding that sSFR is the fundamental physical
parameter which determines the location of massive $z\simeq 1.4$
galaxies on the size-mass plane. 

In light of this apparent contradiction, it is clearly of interest to
ask whether we can reproduce the results of Saracco et al. if we
employ the same criteria to separate the K-Bright
sample. Specifically, if we identify old early-types on the basis of $n\geq
2.5$ and a best-fitting stellar population age $\geq 2$ Gyr, we
find that the K-Bright sample contains twenty-nine such
galaxies. Notably, within this sub-sample of oETGs 22/29 (76\%) would
also be classified as passive according to our adopted sSFR threshold
of $ \leq0.1$ Gyr$^{-1}$. Moreover, if we identify young early-type galaxies in the K-Bright
sample as having $n<2.5$ and age $<2$ Gyr, we find that 7/8 (88\%)
of the galaxies satisfying these criteria would also be classified as
actively star-forming according to our sSFR $>0.1$ Gyr$^{-1}$
criterion. 

Based on the results derived here, we would therefore expect that
separating the Saracco et al. (2009) sample into oETGs and yETGs should 
produce a size-mass plot which looks similar to the right-hand panel of Fig. 5. Specifically, 
we would expect that the oETGs should closely mimic our passive galaxy sub-sample, following a 
size-mass relation with a normalization a factor of $\simeq 2.5$ lower than the local early-type size-mass relation. Furthermore,
we would expect the yETGs to mimic our star-forming sub-sample, lying on the local early-type size-mass relation at $\log(\rm{M}/\Msun)<10.8$, but lying a factor of $\simeq 2$ below it at $\log(\rm{M}/\Msun)\geq10.8$. If we
examine the results  derived by Saracco et al. (e.g. their Fig. 9) it
can be seen that their sub-samples of yETGs and oETGs
display exactly the expected behaviour. 

Consequently, we conclude that the  Saracco et
al. (2009) results are entirely consistent with sSFR being the
fundamental physical parameter governing the location of massive galaxies
on the size-mass plane and that,  amongst morphologically selected
early-type galaxies at $1<z<2$, a stellar population age $\geq 2 $
Gyr is a reasonable, but not one-to-one, proxy for passivity.

\subsubsection{Williams et al. (2010)}
Williams et al. (2010) studied the size evolution of massive galaxies
at $z_{phot}\leq 2$, using a sample of $\simeq 30,000$ ($K<22.4$)
galaxies selected  from the first data release of the UDS. The
stellar-mass estimates derived by Williams et al. were based on the
BC03 template library and assume a Chabrier IMF, again making the
comparison of their results with those derived here straightforward.

In comparison to the Williams et al. sample, the K-Bright sample 
has the advantage of spectroscopic  redshifts,
significantly deeper photometry in the range $1.0\mu$m$-4.5\mu$m and
full spectro-photometric fitting of  the combined photometry and FORS2
spectra. However, in contrast, the Williams et al. sample is $\simeq
350$ times larger than the K-Bright sample and covers a wider
redshift range. 

As part of their investigation, Williams et al. split their sample
into passive and star-forming galaxies using a  sSFR threshold of
sSFR$ = 0.3/t_{H}$, where $t_{H}$ is the age of the Universe at a
given redshift. Notably, over  the redshift range spanned by the
K-Bright sample ($1.3 \leq z \leq 1.5$) this threshold in sSFR is virtually
identical to  the alternative threshold of sSFR$ = 0.1$ Gyr$^{-1}$
adopted here. Over the redshift interval $0.5<z_{phot}<2.0$,  Williams
et al. find that the evolution of the median half-light radii of
massive ($\log({\rm M}/\Msun)\geq 10.8$) passive galaxies evolves
as $\log(R_{c})=0.78(1+z)^{-1.17}$. 

Consequently, at the median redshift of the K-Bright sample, the
results of Williams et al. would predict that the median circularized
half-light  radius of passive K-Bright galaxies with $\log({\rm
  M}/\Msun)\geq 10.8$ should be $1.9\pm 0.1$ kpc.  This is in excellent agreement with
the equivalent value of $R_{c}=1.93\pm{0.24}$ kpc we derive for the passive members of the K-Bright sample.

\subsubsection{van Dokkum et al. (2010)}
In an effort to avoid the effects of progenitor bias, van Dokkum et
al. (2010) used the {\sc newfirm} medium band survey (NMBS)
to select galaxies with a constant number density of $n=2\times
10^{-4}$Mpc$^{-3}$ over the photometric redshift interval $0.2<z<2.2$.
In each of four redshift bins, stacked images of the selected galaxies were 
used to increase the signal-to-noise ratio and allow the evolution of the typical 
galaxy structural parameters to be measured.
Over the redshift interval $0<z<2.2$ van Dokkum et al. (hereafter vD10) found that massive
galaxies selected to have a constant number density of  $n=2\times
10^{-4}$Mpc$^{-3}$ display a factor of $\simeq 4$ growth in their
effective radius, but that the corresponding growth in their stellar
mass is only a factor of $\simeq 2$. As noted by vD10, these results are in agreement
with a galaxy growth model based on minor mergers (see Section 5), a conclusion further supported by 
their finding that the additional stellar-mass growth occurs predominantly at 
large galactic radii (i.e. $r>5$ kpc).

Given that vD10 do not split their sample into
early and late-type morphologies, it is interesting to compare their results with 
our results for the full K-Bright sample, independent of morphological
or sSFR thresholds. Based on their fitting functions for the evolution
of stellar mass, vD10 would predict that galaxies with a constant
number density of $n=2\times 10^{-4}$Mpc$^{-3}$ at $z\simeq1.4$ should
have a median stellar mass of $\log(\rm{M}/\Msun)=11.24$. The  stellar masses for
the vD10 sample were calculated assuming a Kroupa IMF and the Maraston
et al (2005) stellar population models.  If we adopt the conversions
calculated by Cimatti et al. (2008), we see that a stellar mass of
$\log(\rm{M}/\Msun)=11.24$ is equivalent  to $\log(\rm{M}/\Msun)=11.35$ assuming
a Chabrier IMF and the BC03 stellar population models. 
If we then adopt the vD10 policy and define our mass bin as $11.35 \pm {0.15}$, 
we identify a sample of seventeen galaxies within the K-Bright sample. The
corresponding number density of this sub-sample is 
$n=(1.5\pm{0.4})\times 10^{-4}$Mpc$^{-3}$, in good agreement with the vD10 expectation. 
However, our results indicate that the median effective half-light radius
(semi-major axis) of this sub-sample is $3.3\pm{0.3}$ kpc, approximately twenty-five percent smaller
than the vD10 result of $4.3\pm{0.3}$ kpc.

\section{Discussion}
In this final section we provide an extended discussion of the two most important results of this study. 
Firstly, we explore whether any of the commonly proposed evolutionary mechanisms can 
plausibly explain the observed size-mass evolution of the passive members of the K-Bright sample, given the available observational and theoretical constraints. 
Secondly, we discuss the implications for the quenching of star-formation in massive galaxies at high-redshift given our result that a 
substantial fraction of the passive members of the K-Bright sample at $z\simeq 1.4$ have disk-like morphologies.

\subsection{Mechanisms for galaxy size evolution}
Perhaps the cleanest result to emerge from this study is that the passive
(sSFR $\leq 0.1$Gyr$^{-1}$) members of the K-Bright sample follow a
size-mass relation which is identical in slope (and scatter) to that 
followed by early-type galaxies in the SDSS, but off-set to smaller half-light radii 
by a factor of $\simeq 2.4$. As a consequence, in this sub-section we investigate whether any of the
commonly-discussed evolutionary mechanisms are capable of reconciling the size-mass relations for 
passive galaxies at $z\simeq 1.4$ and $z\simeq0$, without violating other observational constraints at low redshift. 

In the following discussion we assume that the
  additional stellar mass added by on-going star-formation amongst the
  passive galaxy sub-sample is negligible. Making the conservative
  assumption that the typical star-formation rate remains constant
  between the observed epoch and the present day, the fractional
  increase in stellar mass from on-going star-formation is $\simeq
  t_{\rm{LB}}\times\langle\rm{sSFR}\rangle$. Taking the median values for
  the passive galaxy sub-sample implies that the fractional increase in
  stellar mass is only $\simeq 18\%$ (0.07 dex).

\subsubsection{AGN feedback}
It has been proposed in the literature that mechanical feedback from
active galactic nuclei (AGN) could be responsible for increasing the
size of massive, compact galaxies at high redshift (e.g. Fan et
al. 2008). In this scenario, AGN driven outflows are
responsible for evacuating material from the central regions of
compact galaxies, resulting in an increase in half-light
radius and a decrease in the central stellar-velocity dispersion. In
fact, because the total mass is conserved, the initial and final values of the 
half-light radius and stellar-velocity dispersion are related as follows:  
$\sigma_{f}/\sigma_{i} = (R_{f}/R_{i})^{-1/2}$.

Using the dynamical mass measurements for the K-Bright galaxies derived here, it is not
possible to rule-out the AGN feedback mechanism as a viable explanation of the observed 
size evolution.  For example, the median stellar mass of the K-Bright sub-sample with
dynamical-mass measurements is $M_{\star}\simeq 10^{11}\Msun$ and the median stellar-velocity dispersion 
is $\sigma_{med}=275$ km s$^{-1}$. Consequently, if their size growth is assumed to be due to AGN feedback, 
the prediction would be that SDSS galaxies with stellar masses of $M_{\star}\simeq 10^{11}\Msun$ should have 
stellar-velocity dispersions of $\sigma \simeq 180$ kms$^{-1}$. This is consistent with what is actually 
observed for early-type galaxies in the SDSS (e.g. van Dokkum, Kriek \& Franx 2009).

However, there are several features of the results derived here which suggest
that AGN feedback is not the dominant mechanism responsible for the size evolution 
of compact high-redshift galaxies. The first is our finding that at $z\simeq 1.4$ 
the off-set from the local size-mass relation is not a function of stellar population age. 
The second is that we find no indication that the scatter in the galaxy size-mass relation is 
significantly greater at $z\simeq 1.4$ than it is locally. Both of these results run counter to the 
expectations of the AGN feedback model as already highlighted by Trujillo et al. (2011), based on 
their study of the size-mass relation of massive early-type galaxies at $z\simeq 1$. Moreover, recent 
numerical simulations suggest that the greatest impact of AGN feedback is likely to occur in much 
younger stellar populations than the $\geq 1$ Gyr old populations seen in massive galaxies at $z\simeq 1.5$ 
(Ragone-Figueroa \& Granato 2011). Finally, it would also appear unlikely that AGN feedback activity can play 
an important role in compact galaxies with low sSFR, given that these galaxies presumably no longer contain a 
large reservoir of gas.

\subsubsection{Major mergers}
Ever since the initial evidence began to emerge that a population of
seemingly-old, highly-compact, early-type galaxies existed at $z\geq
1$, the prospect of explaining their subsequent evolution via a
sequence of major dry (i.e. dissipationless) mergers has been widely
discussed in the literature.  This scenario is immediately appealing
because the product of a major (i.e. mass ratio $\geq$1:3) dry
merger is expected to have an early-type morphology with a half-light
radius which has increased proportionally to the increase in stellar
mass (e.g. Naab et al. 2009). In Fig. 11 the green line illustrates the evolutionary 
path which would be followed by a member of the K-Bright sample if its size evolution is
driven by major mergers.

Although appealing, there are two obvious problems with the major-merger scenario which would 
appear to exclude it as the dominant mechanism responsible for the size evolution of compact high-redshift galaxies. 
Firstly, there is the problem of the
number of major mergers that are actually required to reconcile the
observed size-mass relations at low and high redshift. With reference
to Fig. 11, it can be seen that to shift the locus occupied by the
passive K-Bright galaxies at $z\simeq 1.4$ fully onto the local
relation size-mass relation via major mergers would require a factor of $\simeq 5-6$
increase in stellar mass. This level of major merging is completely inconsistent
with N-body simulations, which suggest that
the typical number of major mergers experienced in the redshift
interval $0<z<1.5$ by galaxies with stellar masses $M_{\star}\geq 10^{11}\Msun$
is $\simeq 1$  (e.g. Hopkins et al. 2010). Moreover, this level of
stellar-mass growth is totally excluded by the constraints provided by
the high-mass end of the local galaxy stellar mass function. The latest derivation of the local galaxy stellar mass
function by  Baldry et al. (2012) suggests that if the passive
members of the K-Bright sample were to grow in stellar mass by a
factor of $\simeq 5$, they  would overshoot the observed number
density of local galaxies with stellar masses $M_{\star}\geq 10^{11.5}\Msun$ by a
factor of $\simeq 25$.

Of course, it could reasonably be argued that the K-Bright sample may
suffer from significant progenitor bias and that the members of the
passive sub-sample of the K-Bright galaxies are not destined to fully
populate the high-mass end of the local early-type size-mass
relation. If, for example, they were only destined to populate the
lower half of the size-mass relation, the amount of stellar-mass growth
required would drop to a factor of $\simeq 2$, seemingly making
the major-merger scenario more plausible. However, if we again compare
with the local galaxy stellar mass function, we find
that the effect of progenitor bias on the K-Bright sample cannot be 
severe. Indeed, the results of Baldry et al. (2012) show 
that the number density of local galaxies with stellar masses $M_{\star}\geq
10^{11}\Msun$ is $n=(3.4\pm{0.4})\times 10^{-4}$ Mpc$^{-3}$. Above this
mass threshold,  where the K-Bright sample should be 100\% complete,
the number density of K-Bright galaxies is $n=(3.0\pm{0.5})\times
10^{-4}$ Mpc$^{-3}$ ($\simeq 90\%$ of the local number density).
Even if we restrict the comparison to only the passive members
of the K-Bright sample, they alone can account for $\simeq 68\%$ of
the local number density of $M_{\star}\geq 10^{11}\Msun$  galaxies.

\begin{figure}
\begin{center}
\includegraphics[scale=0.7]{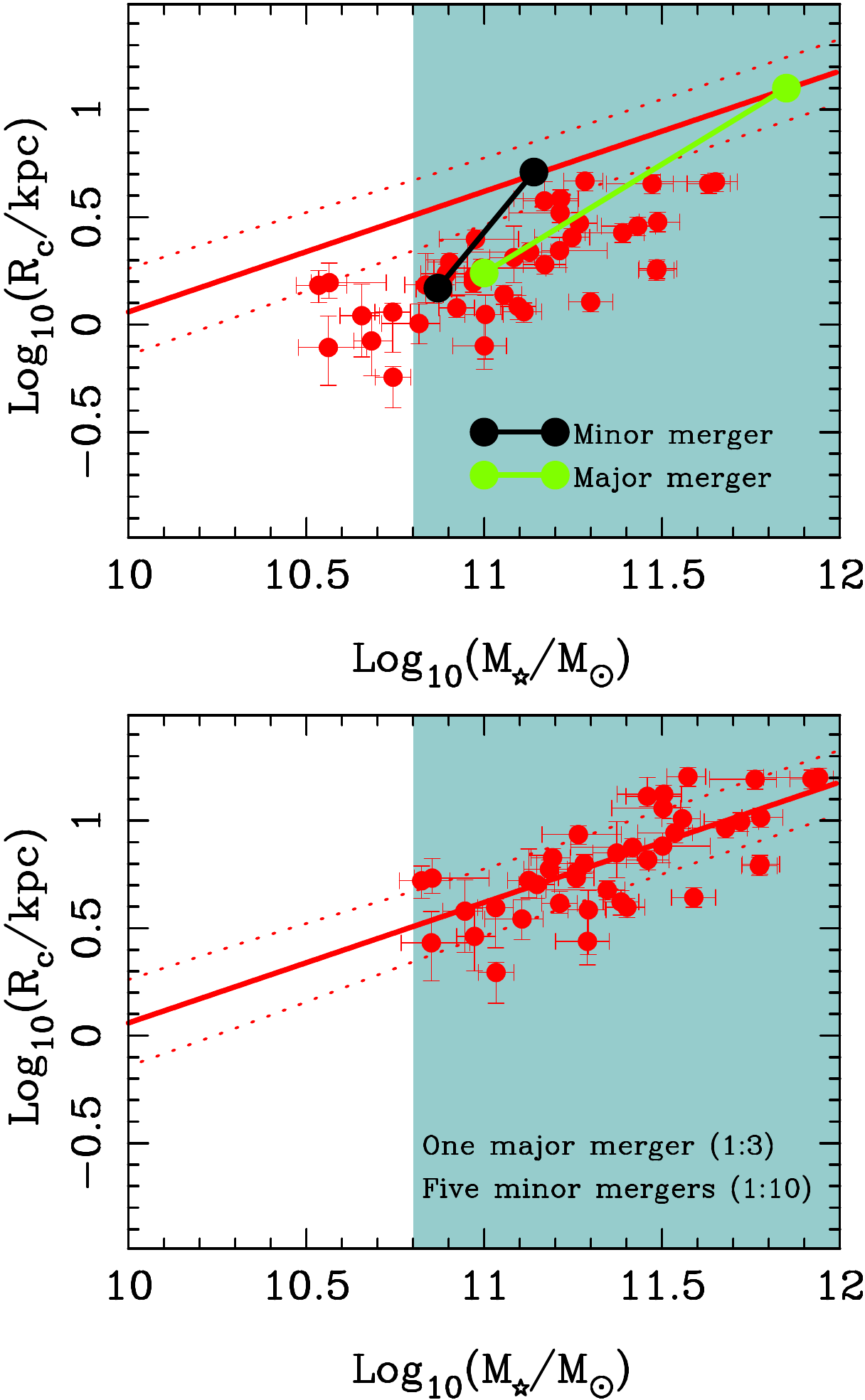}
\caption{The top-panel shows the size-mass relation for the passive members of the
  K-Bright sample, together with the local early-type size-mass relation from
  Shen et al. (2003). The connected black and green points illustrate the evolutionary path 
followed by galaxies if they undergo minor and major mergers respectively. The bottom panel
shows the final location of the passive K-Bright galaxies if they evolve according to our
chosen toy model, which consists of a single major merger (mass ratio 1:3) and five minor (mass ratio 1:10) mergers (see text for a full discussion).}
\end{center}
\end{figure}

\subsubsection{Major plus minor mergers}
Given that there is strong evidence against major mergers being able 
to fully explain the observed size-mass evolution between $z\simeq 1.5$ and the present
day, it is worth considering a toy model in which the additional mass growth is dominated
by minor mergers, which can increase half-light radius in proportion to the square 
of the fractional increase in stellar mass (e.g. Naab et al. 2009). As can be seen from 
the top panel of Fig. 11, this mechanism potentially
allows rapid growth in half-light radius, without requiring stellar-mass growth which
violates the constraints provided by the local galaxy stellar mass function.

In the bottom panel of Fig. 11, we illustrate the expected evolution of the passive members of the K-Bright sample,
based on a toy model which incorporates a mixture of major and minor merging activity. Specifically, the toy model
assumes that, on average, each passive K-Bright galaxy will undergo a single major merger between $z\simeq 1.4$ and the present day and 
that the likely mass ratio of this major merger will be 1:3. 
Additionally, the model also assumes that each galaxy undergoes a series of five minor mergers, each of 
which typically has a mass ratio of 1:10. In this scenario the stellar mass of the galaxy increases by a 
factor of $\simeq 2$ and the half-light radius increases by a factor of $\simeq 3.5$.
As can be seen in the bottom panel of Fig. 11, this toy scenario is at least capable of reconciling the 
observed size-mass relations for massive passive galaxies at $z\simeq 1.4$ and the present day, without 
requiring an unfeasible number of major (or minor) mergers. Moreover, in this toy model, the descendants of 
the passive K-Bright galaxies do not violate the local galaxy stellar mass function because they are 
predicted to contribute $\simeq 95\%$ of local number density of $M_{\star}\geq 10^{11}\Msun$ galaxies.

\subsection{Passive disks and the quenching of star-formation}

As highlighted in Section 4.2, a substantial fraction of the passive members of the K-Bright sample display 
a disk-like morphology. Indeed, of the forty-one members of the K-Bright
sample which are classified as passive, according to our adopted
sSFR$<0.1$Gyr$^{-1}$ criterion, nearly half ($44\%\pm12\%$)
have disk-like morphologies (i.e. $n<2.5$). We note that this is consistent with the
recent results of van der Wel et al. (2011), who estimated that
$65\%\pm 15\%$ of passive galaxies at $z\simeq 2$ with stellar masses
$M_{\star}\geq 6\times 10^{10}\Msun$ have disk-dominated morphologies, based on 
HST imaging from the WFC3/IR Early Release Science (ERS)
programme.  

In contrast, the results of a recent study by Bell et al. (2012), based on S\'{e}rsic profile fitting of the 
CANDELS WFC3/IR imaging in the UKIDSS UDS, suggest that at $z\simeq 1.5$ there are very few passive galaxies 
which do not have a prominent bulge component.
Indeed, Bell et al. (2012) argue that a prominent bulge seems to be a necessary, but not sufficient, condition for quiescence.
Although this would appear to contradict the results derived here, it is clearly possible that the passive disk-like 
galaxies identified in this study could also harbour a significant stellar bulge, which simply cannot be detected 
using ground-based imaging.

However, the high fraction of passive disk-like galaxies at $z\geq 1$ is confirmed by Bruce et al. (2012), 
based on a detailed disk/bulge decomposition analysis of the CANDELS WFC3/IR imaging in the UDS. 
Using a sample of $\simeq 200$ galaxies with $M_{\star}\geq 10^{11}\Msun$ and photometric redshifts in the
range $1.0<z_{phot}<3.0$, Bruce et al. estimate that $25\%\pm6\%$ of
passive (sSFR$<0.1$Gyr$^{-1}$) galaxies are disk-dominated
(i.e. $B/T<0.5$), rising to $40\%\pm7\%$ if they simply classify
disk-dominated objects as having $n<2.5$, based on a single S\'{e}rsic fit.
It is therefore clear that a substantial fraction of massive, passive galaxies at $z>1$ have a
disk-like, or even disk-dominated morphology. 

It is likely that this simple observational fact contains crucial information 
about the evolutionary path followed by the progenitors of the massive galaxy population observed locally. 
As discussed in the previous section, the passive disks in the K-Bright sample are expected to evolve into galaxies 
with $M_{\star}\geq 2\times 10^{11}\Msun$, a mass r\'{e}gime which is dominated 
by early-type galaxies in the local Universe. Consequently, massive disk galaxies at high-redshift which have already been quenched are not 
a natural prediction of evolutionary models in which galaxy-galaxy mergers are responsible for quenching star-formation {\it and} the morphological 
transformation from late-type to early-type.

However, the presence of massive, passive disks at $z\geq 1$ would appear to be in good qualitative 
agreement with recent hydrodynamical simulations of galaxy formation. Such simulations suggest that high-redshift star-formation is dominated 
by filamentary inflows of cold gas ($T<10^{5}$K) until a critical redshift of $z_{crit}\simeq 2$ at which point, at least for the most massive halos, 
accretion of hot halo gas ($T>10^{5}$K) begins to dominate (e.g. Kere\v{s} et al. 2005). 
Moreover, at $z\leq 2$ gas is predicted to be shock heated to $T\simeq T_{vir}$ in halos with masses above a critical threshold ($M_{crit}\simeq 10^{12}\Msun$), suppressing significant further accretion of gas and efficiently 
quenching star formation (e.g. Dekel \& Bimboim 2006). 
A natural prediction of this model is therefore a significant population of massive 
($M_{\star}\geq 10^{10.5}\Msun, \, M_{halo}\geq 10^{12}\Msun$) quenched disks, which are then subsequently 
transformed into spheroids through a combination of both major and minor mergers at lower redshifts.

Finally, we note that this scenario is also in qualitative agreement with the phenomenological model of 
Peng et al. (2010), in which star-formation quenching in the highest-mass galaxies is independent 
of environment and proportional to star-formation rate. This scenario also predicts that morphological transformation and star-formation quenching are separate physical processes and foresees star-forming disks 
rapidly quenching from the blue cloud onto the red sequence.

\section{Conclusions}
In this paper we have presented the results of an analysis of the
relationship between galaxy size,  stellar mass and specific
star-formation rate in a mass complete ($M_{\star}\geq 6\times 10^{11}\Msun$)
sample of spectroscopically confirmed $1.3 \leq z \leq 1.5$ galaxies 
in the UKIDSS UDS field. Using stellar-mass measurements based on
spectro-photometric SED fitting and galaxy size measurements based on
two-dimensional modelling of the available UDS $K-$band imaging, we
have investigated the evolution of massive galaxies on the size-mass
plane as a  function of specific star-formation rate, galaxy
morphology and stellar population age. In addition, we have derived an
alternative constraint on the average size evolution of massive
galaxies since $z\simeq 1.4$ using dynamical mass measurements for a
small sub-sample of galaxies with reliable stellar-velocity dispersion
measurements. Finally, we have investigated evolutionary scenarios which can
plausibly explain the observed size evolution  without violating the
constraints provided by both the latest N-body dark-matter simulations
and current determinations of the local galaxy stellar mass
function. The main conclusions of this study can be summarized as
follows:

\begin{enumerate}

\item{The location of massive galaxies ($M\geq 6\times 10^{10}\Msun$) 
on the size-mass plane at $z\simeq 1.4$ is more closely correlated with sSFR than 
galaxy morphology.}

\item{Massive galaxies at $z\simeq 1.4$ which are passive (sSFR $\leq 0.1$ Gyr$^{-1}$) 
follow a tight size-mass relationship, which is identical in slope (and scatter) 
to that displayed by local early-type galaxies in the SDSS. However, at a given stellar mass, passive galaxies 
at $z\simeq 1.4$ have half-light radii a factor of $f_{g}=2.43\pm{0.20}$ smaller than their low-redshift counterparts.}

\item{In contrast, massive star-forming galaxies at $z\simeq 1.4$
  (sSFR $ >0.1$ Gyr$^{-1}$) lie closer to the size-mass relation of local late-type galaxies, but are still a factor
of $f_{g}=1.61\pm{0.17}$ smaller than their low-redshift counterparts at a given stellar mass.}

\item{If the K-Bright sample is split into early and late-type morphologies using a S\'{e}rsic index
  threshold of $n=2.5$, the corresponding growth factors are found to be: 
  $f_{g}=2.37\pm{0.29}$ and $f_{g}=2.15\pm{0.15}$ for early and late-type galaxies 
respectively. However, the early-type galaxy 
size-mass relation at $z\simeq 1.4$ displays a scatter which is noticeably larger than that associated with the passive 
galaxy size-mass relation. Moreover, the correspondence between passivity and early-type
  morphology (and active star-formation and late-type morphology) in the K-Bright sample is found to be weak.}

\item{Using a small sub-sample of the K-bright galaxies with reliable stellar-velocity 
dispersion measurements, an alternative constraint on the average size evolution
  between $z\simeq 1.4$ and the present day is obtained. The derived value of $f_{g}=2.33\pm{0.32}$ is entirely consistent
  with our primary determination based on stellar-mass measurements.}

\item{We demonstrate that a toy model which combines a single major merger with a sequence of five 
minor mergers can reproduce the observed size evolution 
of massive galaxies, without violating constraints imposed by the local galaxy stellar mass
  function and the predictions of N-body simulations.}

\item{The significant fraction 
of passive galaxies with disk-like morphologies in the K-Bright sample provides additional evidence that 
separate physical processes are responsible for the quenching of
star-formation and the subsequent morphological transformation in massive galaxies at $z\simeq 1.4$.}

\end{enumerate}

\section{acknowledgments}
The authors would like to acknowledge the efforts of the UKIRT staff for observing the 
UKIDSS Ultra-deep survey and St\'{e}phane Charlot for providing the CB07 stellar population models.
RJM would like to acknowledge the funding of the Royal Society via the
award of a University Research Fellowship and the Leverhulme Trust via
the award of a Philip Leverhulme research prize. MC acknowledges the award of an STFC Advanced Fellowship. 
JSD acknowledges the support of the Royal Society via a Wolfson Research Merit award, and
also the support of the European Research Council via the award of an
Advanced Grant. ECL would like to acknowledge financial support from the UK Science
and Technology Facilities Council (STFC) and the Leverhulme Trust. FB acknowledges the support of the 
European Research Council. HJP, RC, and EB acknowledge the award of an STFC PhD studentships. DGB and 
WGH acknowledge funding via the award of STFC PDRA grants. Based in part on data products from observations made 
with ESO Telescopes at the La Silla Paranal Observatory under ESO programme ID 179.A-2006 and on data 
products produced by the Cambridge Astronomy Survey Unit on behalf of the VIDEO consortium.

\bibliographystyle{mn2e}

\begin{appendix}
\section{The K-bright sample}
In Table A1 we provide the basic observational properties and 
important derived quantities for the eighty-one objects in the K-bright sample.

\begin{table*}
\caption{Basic measured and derived properties of the K-bright sample. The first three columns list the catalogue ID number and coordinates 
of each member of the sample. Columns four and five list the spectroscopic redshift (typical uncertainty $\delta z=0.001$) and 
total $K-$band AB magnitude (uncertainty $\leq 5\%$) for each object. Columns six and seven list the derived half-light radii (circularized) and stellar masses along 
with their respective uncertainties. Columns 8-10 list the S\'{e}rsic index (typical uncertainty 30\%), specific star-formation rate and 
stellar population age (typical uncertainty 0.2 dex) for each object respectively. The typical uncertainty on the derived sSFR values is dominated by the 
scatter associated with the adopted star-formation calibrations and is a minimum of a factor of two.}
\begin{tabular}{lcccccccccrc}
\hline
ID & RA(J2000)&DEC(J2000)&$z$&$K_{tot}$&$r_{c}$/kpc & $\log(M_{\star}/\Msun)$ & $n$ & $sSFR/10^{-10}$yr$^{-1}$ &$\log_{10}(age/{\rm yr})$\\
\hline
  34148 &    02:18:38.73 &$-05$:11:41.7& 1.280 &20.5 &$1.2_{-0.2}^{+0.2}$ &    $11.1^{+0.1}_{-0.1}$ &    3.0 &   0.22 &     9.3\\[1ex]
  30370 &    02:18:41.35 &$-05$:13:58.2& 1.289 &21.0 &$2.6_{-0.3}^{+0.3}$ &    $10.8^{+0.1}_{-0.2}$ &    0.9 &             14.5 & 9.6\\[1ex]
  28699 &    02:18:12.60 &$-05$:15:01.2& 1.290 &21.3 &$1.0_{-0.3}^{+0.3}$ &    $10.6^{+0.1}_{-0.1}$ &    3.9 &             18.4 &     9.6\\[1ex]
  69422 &    02:18:43.01 &$-04$:51:17.0& 1.296 &20.3 &$2.2_{-0.4}^{+0.3}$ &    $11.2^{+0.2}_{-0.1}$ &    6.1 &   0.18 &     9.5\\[1ex]
  80073 &    02:18:08.35 &$-04$:45:01.6& 1.299 &20.3 &$3.0_{-0.3}^{+0.4}$ &    $11.3^{+0.1}_{-0.1}$ &    3.6 &   0.02 &     9.5\\[1ex]
  76779 &    02:17:35.88 &$-04$:46:57.1& 1.303 &21.5 &$2.4_{-0.2}^{+0.2}$ &    $10.5^{+0.1}_{-0.1}$ &    1.1 &   1.65 &     8.9\\[1ex]
  65948 &    02:17:27.06 &$-04$:53:18.6& 1.305 &20.7 &$2.5_{-0.3}^{+0.3}$ &    $10.7^{+0.1}_{-0.1}$ &    0.8 &             23.0 &     9.3\\[1ex]
  63940 &    02:16:45.94 &$-04$:54:28.3& 1.311 &21.5 &$1.6_{-0.2}^{+0.4}$ &    $10.6^{+0.2}_{-0.1}$ &    2.8 &   0.08 &     9.4\\[1ex]
  53901 &    02:19:15.37 &$-05$:00:12.8& 1.313 &19.9 &$2.3_{-0.2}^{+0.3}$ &    $11.3^{+0.1}_{-0.1}$ &    3.5 &   1.34 &     9.5\\[1ex]
  32058 &    02:18:47.12 &$-05$:12:56.6& 1.315 &21.1 &$3.7_{-0.9}^{+0.9}$ &    $10.5^{+0.1}_{-0.1}$ &    4.5 &   2.63 &     9.2\\[1ex]
  58266 &    02:19:19.09 &$-04$:57:47.0& 1.316 &20.1 &$2.9_{-0.3}^{+0.3}$ &    $11.3^{+0.1}_{-0.1}$ &    1.1 &             14.1 &     9.6\\[1ex]
  42643 &    02:19:11.84 &$-05$:06:43.8& 1.319 &19.9 &$1.8_{-0.2}^{+0.2}$ &    $11.5^{+0.1}_{-0.1}$ &    3.6 &   0.22 &     9.7\\[1ex]
  79330 &    02:17:19.32 &$-04$:45:22.1& 1.320 &21.2 &$1.2_{-0.2}^{+0.2}$ &    $10.6^{+0.2}_{-0.1}$ &    1.4 &             32.5 &     8.8\\[1ex]
  29501 &    02:18:46.35 &$-05$:14:28.0& 1.320 &21.1 &$3.1_{-0.3}^{+0.3}$ &    $10.6^{+0.1}_{-0.1}$ &    1.0 &   1.48 &     9.1\\[1ex]
  46450 &    02:19:09.47 &$-05$:04:37.0& 1.320 &20.8 &$1.2_{-0.1}^{+0.2}$ &    $10.9^{+0.1}_{-0.1}$ &    1.6 &   0.20 &     9.3\\[1ex]
  77399 &    02:17:04.68 &$-04$:46:31.5& 1.321 &20.1 &$2.9_{-0.3}^{+0.3}$ &    $11.4^{+0.1}_{-0.1}$ &    2.2 &   0.19 &     9.7\\[1ex]
  46091 &    02:19:09.02 &$-05$:04:48.5& 1.321 &20.3 &$2.6_{-0.3}^{+0.3}$ &    $11.3^{+0.1}_{-0.1}$ &    3.9 &   0.04 &     9.6\\[1ex]
  50618 &    02:19:11.78 &$-05$:02:09.2& 1.321 &20.5 &$2.2_{-0.2}^{+0.2}$ &    $11.0^{+0.1}_{-0.1}$ &    2.5 &   1.51 &     9.5\\[1ex]
  65792 &    02:19:14.34 &$-04$:53:23.2& 1.322 &21.1  &$2.5_{-0.3}^{+0.5}$ &    $10.6^{+0.1}_{-0.1}$ &    3.3 &             27.4 &     9.2\\[1ex]
  79998 &    02:17:20.47 &$-04$:45:06.6& 1.323 &21.0 &$1.5_{-0.3}^{+0.6}$ &    $10.8^{+0.1}_{-0.1}$ &    5.3 &   0.61 &     9.5\\[1ex]
  50326 &    02:17:37.02 &$-05$:02:21.7& 1.325 &21.1 &$1.0_{-0.2}^{+0.2}$ &    $10.8^{+0.1}_{-0.1}$ &    4.6 &   0.03  &     9.6\\[1ex]
  77581 &    02:18:01.72 &$-04$:46:27.2& 1.326 &21.1 &$0.8_{-0.3}^{+0.3}$ &    $10.7^{+0.1}_{-0.1}$ &    2.7 &   0.20 &     9.5\\[1ex]
  53783 &    02:17:59.21 &$-05$:00:20.4& 1.329 &21.4 &$0.3_{-0.1}^{+0.1}$ &    $10.5^{+0.1}_{-0.1}$ &    4.1 &   2.20 &     9.2\\[1ex]
  73006 &    02:16:55.04 &$-04$:49:10.4& 1.331 &20.8 &$4.2_{-0.4}^{+0.4}$ &    $10.8^{+0.1}_{-0.1}$ &    1.0 &             13.8 &     9.0\\[1ex]
  60493 &    02:19:19.65 &$-04$:56:25.0& 1.332 &19.8 &$1.8_{-0.2}^{+0.2}$ &    $11.5^{+0.1}_{-0.1}$ &    6.2 &   0.28 &     9.5\\[1ex]
  75483 &    02:16:57.77 &$-04$:47:42.6& 1.372 &20.3 &$3.3_{-0.3}^{+0.3}$ &    $11.2^{+0.1}_{-0.2}$ &    2.5 &   0.33 &     9.6\\[1ex]
  78289 &    02:18:05.26 &$-04$:46:02.9& 1.380 &21.4 &$2.7_{-0.3}^{+0.3}$ &    $10.4^{+0.1}_{-0.1}$ &    1.4 &             14.5 &     9.3\\[1ex]
  73717 &    02:17:22.04 &$-04$:48:46.3& 1.380 &21.3 &$1.4_{-0.3}^{+0.4}$ &    $10.4^{+0.1}_{-0.1}$ &    1.4 &             44.7 &     9.3\\[1ex]
  42941 &    02:17:37.90 &$-05$:06:37.2& 1.394 &21.1 &$1.9_{-0.2}^{+0.2}$ &    $10.8^{+0.2}_{-0.1}$ &    2.0 &   1.25 &     9.2\\[1ex]
  77327 &    02:18:02.96 &$-04$:46:36.4& 1.399 &20.6 &$1.2_{-0.1}^{+0.1}$ &    $11.1^{+0.1}_{-0.1}$ &    3.5 &   0.07 &     9.5\\[1ex]
  79158 &    02:17:58.54 &$-04$:45:29.9& 1.399 &20.0 &$2.7_{-0.3}^{+0.3}$ &    $11.4^{+0.1}_{-0.2}$ &    1.5 &   0.35 &     9.5\\[1ex]
  49095 &    02:18:46.72 &$-05$:03:03.7& 1.401 &20.8 &$3.1_{-0.5}^{+1.2}$ &    $11.0^{+0.1}_{-0.1}$ &    5.8 &   1.30 &     9.5\\[1ex]
  79274 &    02:17:05.57 &$-04$:45:30.3& 1.401 &21.3 &$1.8_{-0.2}^{+0.2}$ &    $10.7^{+0.1}_{-0.1}$ &    1.8 &   3.36 &     9.6\\[1ex]
  45629 &    02:19:10.14 &$-05$:05:07.6& 1.401 &21.4 &$0.6_{-0.2}^{+0.1}$ &    $10.7^{+0.1}_{-0.1}$ &    5.6 &   0.20 &     9.4\\[1ex]  
  61110 &    02:18:32.62 &$-04$:56:04.1& 1.402 &20.6 &$2.5_{-0.3}^{+0.3}$ &    $11.0^{+0.1}_{-0.1}$ &    2.1 &   0.73 &     9.5\\[1ex]
  56208 &    02:16:23.91 &$-04$:58:59.0& 1.402 &20.7 &$1.7_{-0.2}^{+0.4}$ &    $10.9^{+0.1}_{-0.1}$ &    4.9 &   1.09 &     9.3\\[1ex]
  54647 &    02:18:49.81 &$-04$:59:51.2& 1.404 &21.1 &$1.1_{-0.4}^{+0.3}$ &    $11.0^{+0.1}_{-0.1}$ &    8.0 &   0.50 &     9.6\\[1ex]
  63911 &    02:16:17.68 &$-04$:54:26.8& 1.405 &20.5 &$3.8_{-0.4}^{+0.9}$ &    $11.2^{+0.1}_{-0.1}$ &    3.5 &   0.04 &     9.5\\[1ex]
  57918 &    02:18:34.48 &$-04$:58:00.1& 1.407 &20.8 &$1.4_{-0.1}^{+0.1}$ &    $11.1^{+0.1}_{-0.2}$ &    1.6 &   0.02 &     9.6\\[1ex]
  56439 &    02:17:23.07 &$-04$:58:47.8& 1.408 &19.8 &$4.5_{-0.5}^{+0.5}$ &    $11.6^{+0.1}_{-0.1}$ &    2.0 &   0.37 &     9.6\\[1ex]
  45372 &    02:18:31.67 &$-05$:05:14.8& 1.408 &20.4 &$3.9_{-0.4}^{+0.4}$ &    $11.2^{+0.1}_{-0.2}$ &    2.3 &   0.54 &     9.5\\[1ex]
\hline
\end{tabular}
\end{table*}

\setcounter{table}{0}
\begin{table*}
\caption{Continued.}
\begin{tabular}{lcccccccccrc}
\hline
ID & RA(J2000)&DEC(J2000)&$z$&$K_{tot}$&$r_{c}$/kpc & $\log(M_{\star}/\Msun)$ & $n$ & $sSFR/10^{-10}$yr$^{-1}$ &$\log(age/{\rm yr})$\\
\hline
  44194 &    02:19:25.35 &$-05$:05:52.1& 1.408 &20.2 &$2.4_{-0.2}^{+0.2}$ &    $11.0^{+0.1}_{-0.1}$ &    1.6 &   2.98 &     9.2\\[1ex]
  54522 &    02:17:32.14 &$-04$:59:55.0& 1.411 &21.0 &$1.5_{-0.2}^{+0.5}$ &    $10.9^{+0.1}_{-0.1}$ &    3.7 &   0.05 &     9.3\\[1ex]
  59320 &    02:16:06.65 &$-04$:57:06.8& 1.409 &20.0 &$4.5_{-0.5}^{+0.5}$ &    $11.5^{+0.1}_{-0.2}$ &    2.2 &   0.36 &     9.4\\[1ex]
  32227 &    02:16:59.40 &$-05$:12:50.7& 1.410 &20.9 &$2.3_{-0.2}^{+0.2}$ &    $10.9^{+0.2}_{-0.1}$ &    1.0 &   19.9 &     9.1\\[1ex]
  49961 &    02:17:36.52 &$-05$:02:33.0& 1.411 &20.5 &$1.9_{-0.2}^{+0.2}$ &    $11.2^{+0.1}_{-0.1}$ &    2.9 &   0.48 &     9.5\\[1ex]
  63237 &    02:18:07.67 &$-04$:54:51.4& 1.412 &21.0 &$2.0_{-0.2}^{+0.2}$ &    $10.9^{+0.1}_{-0.1}$ &    2.1 &   0.23 &     9.5\\[1ex]
  46492 &    02:18:38.78 &$-05$:04:34.2& 1.414 &20.5 &$3.5_{-0.4}^{+0.4}$ &    $11.0^{+0.1}_{-0.1}$ &    1.2 &             15.0  &     9.4\\[1ex]
  52436 &    02:17:36.41 &$-05$:01:07.1& 1.421 &20.5 &$1.1_{-0.1}^{+0.1}$ &    $11.1^{+0.1}_{-0.1}$ &    2.6 &             21.5 &     9.3\\[1ex]
  63675 &    02:16:52.34 &$-04$:54:37.3& 1.429 &21.1 &$1.7_{-0.2}^{+0.2}$ &    $10.8^{+0.1}_{-0.1}$ &    1.1 &             20.2 &     9.0\\[1ex]
  81348 &    02:17:02.98 &$-04$:44:21.3& 1.435 &21.5 &$2.7_{-0.3}^{+0.3}$ &    $10.7^{+0.1}_{-0.1}$ &    1.2 &   9.40 &     9.6\\[1ex]
  29788 &    02:16:59.29 &$-05$:14:19.9& 1.435 &21.4 &$1.1_{-0.4}^{+0.4}$ &    $10.7^{+0.1}_{-0.1}$ &    1.9 &   0.63 &     9.3\\[1ex]
  62125 &    02:17:18.68 &$-04$:55:26.7& 1.441 &19.8 &$4.6_{-0.5}^{+0.5}$ &    $11.7^{+0.1}_{-0.2}$ &    5.6 &   0.02  &     9.6\\[1ex]
  53841 &    02:17:02.42 &$-05$:00:18.0& 1.444 &20.8 &$1.8_{-0.2}^{+0.2}$ &    $11.0^{+0.2}_{-0.1}$ &    1.8 &   0.22 &     9.2\\[1ex]
  46886 &    02:17:05.81 &$-05$:04:23.1& 1.451 &21.3 &$1.5_{-0.3}^{+0.3}$ &    $10.5^{+0.1}_{-0.1}$ &    1.3 &   0.93 &     9.1\\[1ex]
  78217 &    02:17:21.67 &$-04$:46:03.4& 1.456 &20.7 &$3.0_{-0.3}^{+0.3}$ &    $11.0^{+0.1}_{-0.1}$ &    3.5 &   1.00  &     9.5\\[1ex]
  48550 &    02:18:06.02 &$-05$:03:26.1& 1.456 &20.9 &$1.7_{-0.2}^{+0.2}$ &    $11.0^{+0.1}_{-0.2}$ &    1.3 &   7.68 &     9.4\\[1ex]
  47359 &    02:18:42.06 &$-05$:03:58.5& 1.456 &20.3 &$4.7_{-0.5}^{+0.5}$ &    $11.3^{+0.1}_{-0.1}$ &    2.5 &   0.08 &     9.5\\[1ex]
  73600 &    02:17:06.30 &$-04$:48:50.4& 1.458 &21.2 &$1.1_{-0.4}^{+0.1}$ &    $10.7^{+0.1}_{-0.1}$ &    0.9 &   0.06 &     9.3\\[1ex]
  58689 &    02:17:19.26 &$-04$:57:34.1& 1.459 &21.3 &$4.5_{-1.8}^{+0.5}$ &    $10.8^{+0.1}_{-0.1}$ &    8.0 &   1.15 &     9.2\\[1ex]
  64357 &    02:17:17.46 &$-04$:54:13.5& 1.460 &21.3 &$2.6_{-0.3}^{+0.3}$ &    $10.6^{+0.1}_{-0.1}$ &    1.0 &             19.7 &     9.4\\[1ex]
  79138 &    02:17:20.51 &$-04$:45:32.4& 1.461 &20.9 &$2.8_{-1.1}^{+0.3}$ &    $10.8^{+0.1}_{-0.1}$ &    2.5 &             23.4 &     9.2\\[1ex]
  78923 &    02:17:20.52 &$-04$:45:41.3& 1.462 &20.8 &$0.8_{-0.2}^{+0.3}$ &    $11.0^{+0.1}_{-0.1}$ &    6.0 &   0.14 &     9.6\\[1ex]
  44334 &    02:17:24.56 &$-05$:05:48.6& 1.467 &20.1 &$3.0_{-0.3}^{+0.3}$ &    $11.4^{+0.1}_{-0.1}$ &    2.5 &   5.09 &     9.5\\[1ex]
  62775 &    02:16:48.66 &$-04$:55:06.2& 1.467 &21.1 &$1.4_{-0.4}^{+0.3}$ &    $10.9^{+0.2}_{-0.1}$ &    1.1 &   5.73 &     9.2\\[1ex]
  66424 &    02:17:00.91 &$-04$:53:03.4& 1.467 &20.9 &$2.2_{-0.2}^{+0.2}$ &    $11.1^{+0.1}_{-0.2}$ &    1.5 &   0.80 &     9.5\\[1ex]
  71384 &    02:17:25.54 &$-04$:50:07.9& 1.477 &20.0 &$3.0_{-0.3}^{+0.3}$ &    $11.5^{+0.1}_{-0.1}$ &    3.5 &   0.26 &     9.6\\[1ex]
  72088 &    02:17:18.88 &$-04$:49:45.7& 1.477 &21.4 &$1.2_{-0.4}^{+0.5}$ &    $10.8^{+0.1}_{-0.2}$ &    2.1 &   2.04 &     9.5\\[1ex]
  72815 &    02:17:30.38 &$-04$:49:18.2& 1.477 &21.3 &$3.0_{-0.3}^{+0.3}$ &    $10.6^{+0.1}_{-0.1}$ &    0.5 &             32.4&     8.9\\[1ex]
  70067 &    02:17:24.38 &$-04$:50:55.5& 1.478 &21.1 &$2.1_{-0.5}^{+0.8}$ &    $11.1^{+0.1}_{-0.1}$ &    5.9 &   0.27 &     9.5\\[1ex]
  60843 &    02:18:40.11 &$-04$:56:14.5& 1.478 &21.4 &$3.9_{-0.4}^{+0.4}$ &    $10.6^{+0.1}_{-0.2}$ &    0.8 &             35.3 &     9.2\\[1ex]
  53230 &    02:18:50.65 &$-05$:00:36.3& 1.478 &20.6 &$1.9_{-0.5}^{+0.2}$ &    $11.1^{+0.1}_{-0.1}$ &    8.0 &   5.13 &     9.4\\[1ex]
  56151 &    02:18:43.61 &$-04$:59:01.1& 1.483 &20.8 &$1.6_{-0.2}^{+0.2}$ &    $11.0^{+0.1}_{-0.1}$ &    1.7 &   0.44 &     9.4\\[1ex]
  52354 &    02:16:52.72 &$-05$:01:11.0& 1.483 &21.4 &$3.1_{-0.3}^{+0.3}$ &    $10.5^{+0.1}_{-0.1}$ &    0.9 &             42.5 &     9.0\\[1ex]
  29201 &    02:17:19.95 &$-05$:14:40.5& 1.485 &21.4 &$0.8_{-0.3}^{+0.3}$ &    $10.6^{+0.1}_{-0.1}$ &    3.5 &   0.75 &     9.1\\[1ex]
  48451 &    02:19:13.83 &$-05$:03:25.4& 1.485 &20.3 &$3.5_{-0.4}^{+0.4}$ &    $11.2^{+0.1}_{-0.1}$ &    2.2 &             11.0 &     9.3\\[1ex]
  43168 &    02:18:33.52 &$-05$:06:28.1& 1.491 &21.0 &$1.4_{-0.6}^{+0.2}$ &    $10.9^{+0.1}_{-0.2}$ &    7.7 &   1.84 &     9.4\\[1ex]
  47774 &    02:18:30.39 &$-05$:03:50.1& 1.497 &21.1 &$3.6_{-0.4}^{+0.8}$ &    $10.9^{+0.1}_{-0.1}$ &    2.5 &   2.93 &     9.4\\[1ex]
  50229 &    02:18:38.09 &$-05$:02:24.3& 1.498 &21.0 &$1.7_{-0.2}^{+0.2}$ &    $10.9^{+0.1}_{-0.1}$ &    1.5 &   0.16 &     9.5\\[1ex]
  54253 &    02:16:22.61 &$-05$:00:01.8& 1.502 &20.4 &$1.3_{-0.1}^{+0.1}$ &    $11.3^{+0.1}_{-0.1}$ &    4.1 &   0.04 &     9.5\\[1ex]
  61727 &    02:18:02.04 &$-04$:55:43.6& 1.505 &20.7 &$1.7_{-0.2}^{+0.2}$ &    $11.0^{+0.1}_{-0.2}$ &    2.4 &   0.02 &     9.4\\[1ex]
\hline
\end{tabular}
\end{table*}

\end{appendix}

\end{document}